\documentclass[11pt]{article}
\usepackage{jheppub}
\usepackage{amsmath,amssymb,amsfonts,graphicx,slashed,amsthm,mathtools,upgreek, enumerate, tensor, subfig}
\usepackage[dvipsnames]{xcolor}
\usepackage{arydshln}

\usepackage{comment}
\usepackage{hyperref}
\usepackage[utf8]{inputenc}
\usepackage[titletoc]{appendix}

\numberwithin{equation}{section} 
\allowdisplaybreaks

\allowdisplaybreaks

\newcommand{\be}{\begin{equation}}
\newcommand{\ee}{\end{equation}}
\newcommand{\f}{\frac}
\newcommand{\s}{\sqrt}
\newcommand{\p}{\partial}

\newcommand{\bea}{\begin{eqnarray}}
\newcommand{\eea}{\end{eqnarray}}
\newcommand{\ba}{\begin{align}}
\newcommand{\ea}{\end{align}}

\newcommand{\la}{\langle}
\newcommand{\ra}{\rangle}
\newcommand{\beq}{\begin{equation}}
\newcommand{\eeq}{\end{equation}}

\newcommand{\ket}[1]{| #1 \rangle}

\newcommand{\avg}[1]{\langle #1 \rangle}

\DeclareMathOperator{\tr}{tr}

\title{Islands in de Sitter space}

\author[a, b]{Vijay Balasubramanian}
\author[a]{\!, Arjun Kar}
\author[c,d,a]{\!, Tomonori Ugajin}

\affiliation[\,a]{David Rittenhouse Laboratory, University of Pennsylvania,\\
209 S.33rd Street, Philadelphia, PA 19104, USA}
\affiliation[\,b]{Theoretische Natuurkunde, Vrije Universiteit Brussel (VUB), and \\ International Solvay Institutes, Pleinlaan 2, B-1050 Brussels, Belgium}
\affiliation[\,c]{Center for Gravitational Physics,
Yukawa Institute for Theoretical Physics, Kyoto University,\\
Kitashirakawa Oiwakecho, Sakyo-ku,
Kyoto 606-8502, Japan}
\affiliation[\,d]{The Hakubi Center for Advanced Research, Kyoto University,\\
Yoshida Ushinomiyacho, Sakyo-ku, Kyoto 606-8501, Japan}
\emailAdd{vijay@physics.upenn.edu}
\emailAdd{arjunkar@sas.upenn.edu}
\emailAdd{tomonori.ugajin@yukawa.kyoto-u.ac.jp}

\abstract{
We consider black holes in 2d de Sitter JT gravity coupled to a CFT, and entangled with matter in a disjoint non-gravitating universe. Tracing out the entangling matter leaves the CFT in a density matrix whose stress tensor backreacts on the de Sitter geometry, lengthening the wormhole behind the black hole horizon. Naively, the entropy of the entangling matter increases without bound as the strength of the entanglement increases, but the monogamy property predicts that this growth must level off.   We compute the entropy via the replica trick, including  wormholes between the replica copies of the de Sitter geometry, and find  a competition between conventional field theory entanglement entropy and the surface area of extremal ``islands'' in the de Sitter geometry.  The black hole and cosmological horizons both play a role in generating such islands in the back-reacted  geometry, and have the effect of stabilizing the entropy growth as required by monogamy.   We first show this in a scenario in which the de Sitter spatial section has been decompactified to an interval.  Then we consider the compact geometry, and argue for a novel interpretation of the island formula in the context of closed universes that recovers the Page curve.  Finally, we comment on the application of our construction to the cosmological horizon in empty de Sitter space.
}


\begin{document}

\maketitle

\parskip=10pt

\section{Introduction}

Extremal islands are gravitating regions which are reconstructible from quantum information stored in  entangled non-gravitating system. These effects are captured succinctly by the so-called island formula \cite{Almheiri:2019hni}, which computes the fine-grained entanglement entropy of the non-gravitating system. This formula is closely related to the holographic formula for entanglement entropy \cite{Ryu:2006bv,Ryu:2006ef,Hubeny:2007xt} and its quantum corrections \cite{Faulkner:2013ana, Engelhardt:2014gca,Almheiri:2019psf, Penington:2019npb}.  In the derivation of this formula, the naive effective field theory calculation of the radiation entropy is corrected by a non-perturbative gravitational effect: Euclidean wormholes which appear in the replica trick \cite{Almheiri:2019qdq,Penington:2019kki}.
Such wormholes provide additional saddle-points for the gravitational path integral used to compute R\'enyi entropies, and these saddles dominate the entropy calculation after the Page time.  In black hole evaporation, their net effect is to halt the linear growth of the entropy predicted by Hawking's calculation \cite{Hawking:1974sw}.  This truncation is required by unitarity because the dimension of the Hilbert space of black hole microstates is finite.  
The island formula has been also successfully applied to asymptotically flat black holes \cite{Anegawa:2020ezn, Hashimoto:2020cas, Gautason:2020tmk,Hartman:2020swn,Dong:2020uxp,Krishnan:2020oun}, higher dimensions \cite{Almheiri:2019psy,Balasubramanian:2020hfs}, and cosmology \cite{Dong:2020uxp,Chen:2020tes,Hartman:2020khs,VanRaamsdonk:2020tlr}.\footnote{For further work on the island formula in general contexts, see \cite{Almheiri:2020cfm} and references therein.  These developments have also led to a revival of interest in baby universes and ensemble interpretations of gravity \cite{Saad:2019lba,Marolf:2020xie} (see \cite{Balasubramanian:2020jhl} for more complete references and a discussion of the main concepts in a simple model).}

Following \cite{Penington:2019kki}, consider two disjoint universes, $A$ and $B$.   We place quantum matter in each universe so that the total Hilbert space is naturally bipartite, $\mathcal{H}_{\text{tot}} = \mathcal{H}_A \otimes \mathcal{H}_B$.  
In this paper, we work in the semiclassical limit where the spatial topologies of the universes do not undergo quantum fluctuations.\footnote{These effects can be important in certain situations involving baby universes, along the lines of \cite{Marolf:2020xie,Anous:2020lka,Chen:2020tes}. }
Since the two universes are disconnected, classical exchange of information is prohibited. However, as in \cite{Penington:2019kki}, we can still consider entangled quantum states. Specifically, we consider some purification of a thermal state, e.g., the thermofield double (TFD) state, in which one can by tune the amount of entanglement by  changing the TFD temperature $1/\beta$. 
We now turn on semiclassical JT-de Sitter gravity  on universe $B$.    
In \cite{WIP}, using the replica wormhole argument, an island formula was derived for this situation, and for the case with a negative cosmological constant in universe $B$. In the latter case, backreaction of the stress tensor caused by the increasing temperature created a causal shadow region behind the horizon, and this region was identified with the island. As a result, it was shown that the entanglement entropy between two disjoint universes, one with AdS gravity, follows the  Page curve (i.e., the linear growth and subsequent saturation) as we increase the temperature.   In this paper we apply a similar procedure to two dimensional Jackiw-Teitelboim (JT) gravity with a positive  cosmological constant.

One of the mysteries of de Sitter space is that it has an entropy associated with a ``cosmological horizon".
It was argued that this entropy implies the Hilbert space dimension of de Sitter quantum gravity is finite  \cite{Banks:2000fe}.\footnote{See \cite{Bousso:2002fq} for a nice review of de Sitter space and its quantum aspects.} This seems to contradict the fact that the dimension of the Hilbert space of effective field theory in de Sitter space is infinite. This tension is similar in spirit to the black hole information paradox, in the sense that it appears impossible for the gravitational Hilbert space to accommodate all of the effective field theory degrees of freedom. Though we will certainly not resolve this issue, we will be able to show that when de Sitter degrees of freedom become entangled with another universe, an extremal island develops near the cosmological horizon. Our result suggests that the island formula, which is derived within the framework of semiclassical gravity coupled to effective field theory, can detect the finite dimensionality of the de Sitter quantum gravity Hilbert space.

We can also include a black hole in our model in order to study de Sitter black hole evaporation.  Indeed, two dimensional de Sitter JT gravity is obtained from the dimensional reduction of the Nariai limit of higher  dimensional Schwarzchild-de Sitter black holes, where by extremal we mean the mass of the black hole becomes maximal.
In the Nariai limit, the cosmological horizon and the black hole horizon are in  thermal equilibrium since they have the same temperature. 
However, as shown in 2d cases \cite{Bousso:1997wi, Nojiri:1998ue,Nojiri:1998ph}, near-extremal de Sitter black holes do indeed evaporate and have empty de Sitter as an endpoint of their evolution. We also refer readers to \cite{Ginsparg:1982rs} for a study of the classical dynamics of the Nariai metric. In the presence of the black hole, we find that there are several possible qualitatively distinct islands. The appearance of these islands is directly associated with the existence of both the black hole horizon and cosmological horizon in this spacetime, and  subtle global considerations are necessary to observe the expected Page-like behavior for the entropy of universe $A$.

\subsection{Can auxiliary systems be entangled with closed universes?}

The island formula instructs us to minimize and extremize the generalized entropy, i.e. the surface area of the island plus effective field theory entropy of the island plus the auxiliary system, over all possible subregions of the gravitating universe.  Therefore, taking the entire Cauchy slice of the closed universe as the island will yield a generalized entropy of zero, since the total system is in a pure state and the closed universe has no boundary with which to contribute an area term.  This suggests that we cannot construct pure entangled states of a closed gravitating universe and an auxiliary systems (in \cite{Almheiri:2019hni}, a single qubit was used to illustrate the point).

Let us try to understand this issue by considering various interpretations of the island formula. We have an auxiliary system with Hilbert space $\mathcal{H}_A$ and a closed universe with a quantum gravity Hilbert space $\mathcal{H}_{G}$. By the axioms of quantum mechanics, the total Hilbert space is $\mathcal{H}_{AG} = \mathcal{H}_A \otimes \mathcal{H}_G$. So far, we have used only axioms and the assumption that quantum gravity in a closed universe has a space of states with the structure of a Hilbert space. One interpretation of the general argument (outlined above) involving the island formula in this situation appears to imply that there are no entangled states in $\mathcal{H}_{AG}$.
However, if $\dim \mathcal{H}_G > 1$ (we can always choose $\dim \mathcal{H}_A > 1$), we can pick two of these disentangled states which obey the implication of the island formula. Call them $\ket{0}_A \ket{0}_G$ and $\ket{1}_A\ket{1}_G$.
Since $\mathcal{H}_{AG}$ is a Hilbert space, it is closed under addition.
This means we can construct the Bell pair $\ket{0}_A\ket{0}_G + \ket{1}_A\ket{1}_G$, which is clearly entangled.  But the strong application of the island formula would say that no such entangled states are possible, and would lead us to conclude that $\dim \mathcal{H}_G = 1$. Alternatively, there would have to be subtle gravitational constraints which modify the axioms we have used in this argument in an unknown way which only appears in the discussion of closed universes. Both of these options are clearly in tension with  general beliefs, for example, that the Hilbert space of de Sitter quantum gravity has dimension of order $e^{1/G_N}$ \cite{Banks:2000fe,Witten:2001kn}.\footnote{But see, for instance, \cite{McNamara:2020uza} where it is conjectured that in fact $\dim \mathcal{H}_G = 1$ from swampland reasoning, and also \cite{Hsin:2020mfa} where the absence of global symmetries are used to argue for this conclusion. }

Perhaps a less dramatic interpretation of the general argument in \cite{Almheiri:2019hni} is that the island formula only applies to semiclassical states in $\mathcal{H}_G$, and it is these states which cannot be entangled with auxiliary systems.  However, there are mixed states $\rho_G$ on $\mathcal{H}_G$ which have nonzero entropy, and if $\mathcal{H}_G$ has black hole microstates (as is believed for de Sitter) then we expect a generic highly entangled mixed state constructed from these to correspond to a semiclassical black hole background.
But then, by general principles, we can always purify such a $\rho_G$ by using an auxiliary system $\mathcal{H}_A$, yielding a pure state $\ket{\psi}_{AG} \in \mathcal{H}_{AG}$ which is clearly entangled with entropy $S(\rho_G) \neq 0$.
Indeed, we would instead have $S(\rho_G) = S_{BH}$.

We will argue that these conceptual difficulties can be avoided by adopting an  alternative interpretation of the island formula which can be applied here.
Instead of taking the entire Cauchy slice as the island in a closed universe, we instead consider the Cauchy slice minus a puncture (understood as the limit of the Cauchy surface minus a small sphere surrounding the puncture). The area contribution from the small sphere will be large since $1/G_N$ is large, and we will see that limit of the entropy as the sphere radius vanishes can be finite.  Thus, in the weak entanglement regime, the island formula interpreted this way would lead us to  conclude that the entanglement of the auxiliary system with the closed universe is just the naive field theory entropy. Then, when the   entanglement  becomes sufficiently large, the area of the small sphere dominates the entropy calculation.  In this way, it is possible for a auxiliary system to to be entangled with a closed, gravitating universe while respecting the Page behavior.  We will see that precisely this scenario occurs in the de Sitter black hole with the punctures in question being located at the apparent horizon.

We will present several justifications of our proposed prescription.  One clear approach is to decompatify the de Sitter circle by considering the universal cover of the de Sitter black hole geometry.  In this case, the universe is not compact and we will show that the island story proceeds in a more or less standard way, and shows Page-like behavior consistent with entanglement monogamy.   We expect the compact situation to yield similar results because the island formula involve local extremization conditions.   In addition, there is evidence from AdS/CFT that maximally extending a geometry does not involve changing the underlying quantum theory or state \cite{Balasubramanian:2019qwk}. Indeed, we re-compactify the extended geometry by introducing identifications on the spatial slice, recovering our proposed prescription for islands on compact universes.

\subsection{Outline}

In Sec.~\ref{sec:setup}, we set up the entangled system by choosing a state on two disjoint universes, and we review the specific gravitational and quantum matter theories  which we study. We then briefly review the main results of \cite{WIP}, in particular the statement that the entanglement entropy between the two universes is given by a generalized entropy on the gravitating universe $B$. In Sec.~\ref{sec:semiclassical}, we give a semiclassical solution of the 2d de Sitter JT equations of motion with backreaction from thermal matter fields. In Sec.~\ref{sec:penrose}, we study the effect of this backreaction on the Penrose diagram of the 2d de Sitter black hole.  In particular, we show that it develops a long interior region as we increase the entanglement between two universes. 
In Sec.~\ref{sec:island}, we study the generalized entropy of the gravitating universe, and derive a Page-like curve for the black hole background and comment on the pure de Sitter situation.

As this work was nearing completion, complementary work on semiclassical gravitational entropy in cosmological spacetimes appeared in \cite{Chen:2020tes,Hartman:2020khs,VanRaamsdonk:2020tlr}.\footnote{The authors of \cite{Chen:2020tes} formulated an intriguing paradox involving subadditivity, the resolution of which required inclusion of bra-ket wormholes.  Their discussion involved Lorentzian de Sitter evolution; it would be interesting to understand if on-shell Euclidean bra-ket wormholes play a role in resolving some of the issues with the Euclidean de Sitter construction that we have pointed out in this paper.}

\section{Setup}\label{sec:setup}

\subsection{Review of previous work}

We begin by reviewing the setup in \cite{WIP}, adapted for application to a de Sitter universe. We are interested in islands in two dimensional de Sitter space which emerge when the de Sitter degrees of freedom are entangled with fields living in another universe.   To this end, we prepare two disjoint universes, $A$ and $B$. Each universe supports a conformal quantum field theory, and for convenience we choose the same CFT on both. Importantly, we turn on gravity ($G_N \neq 0$) only on universe $B$, and assume it to be asymptotically de Sitter space. By contrast, universe $A$ (though it may have curvature) has $G_N = 0$.
To summarize, the effective action of each universe is given by 
\be 
\log Z_{A} = \log Z_{\text{CFT}}[A], \quad \log Z_{B} = -I_{\text{grav}} +\log Z_{\text{CFT}}[B]  .
\ee
In addition, we choose  $I_{\text{grav}} $ to be the action of de Sitter JT gravity.   We treat the gravitational sector on universe $B$ semiclassically.  The total Hilbert space of quantum fields in this system is naturally bipartite, $\mathcal{H}_{\text{tot}} = \mathcal{H}_{A} \otimes \mathcal{H}_{B}$, where $\mathcal{H}_{A}$ and $\mathcal{H}_{B}$ are the Hilbert spaces of the CFT on $A$ and $B$, respectively.
Using a formula for microscopic entropy studied in \cite{WIP}, we will study the entanglement structure of the following state in the presence of gravity on universe $B$,
\be 
| \Psi \ra = \sum_{i} ^{\infty}  \s{p_{i}} | i \ra_{A} \otimes | \psi_{i} \ra_{B}, \quad  p_{i}=\f{e^{-\beta E_{i}}}{Z_{\beta}} , \label{eq:HHstate}
\ee
where $\{ | i \ra_{A}\} $ are an orthogonal basis of $\mathcal{H}_{A} $ and  $|\psi_{i} \ra_{B}$ is an energy eigenstate of the CFT on universe $B$ with eigenvalues $E_{i}$.
Though it is not strictly necessary, we could choose $\ket{i}_A$ to also be energy eigenstates if we wish.
We also defined the partition function $Z_{\beta} =\sum_{i} e^{-\beta E_{i}}$.
The parameter $\beta$ (the inverse CFT temperature) in the definition of the state \eqref{eq:HHstate} controls the amount of the entanglement between the two universes.

In \cite{WIP}, the entanglement entropy of the state \eqref{eq:HHstate} was computed for the non-gravitating universe $A$ using the replica trick.
Specifically, we are interested in the reduced density matrix of \eqref{eq:HHstate} on $A$
\be
\rho_{A} =\sum_{i,j=1}^{\infty} \s{p_{i} p_{j}}\; \la \psi_{i} | \psi_{j} \ra_{B} \; | i \ra \la j |_A , 
\ee
and the $n \rightarrow 1$ limit of the R\'enyi entropy 
\be 
\tr \rho_{A}^{n} = \sum_{i_{1} \cdots i_{n}} p_{i_{1}} \cdots p_{i_{n}}  \la \psi_{i_{1}} |\psi_{i_{2}} \ra \cdots \la  \psi_{i_{n}} |\psi_{i_{1}} \ra .
\ee
The computation of the right hand side of the above R\'enyi entropy involves the gravitational path integral on $n$ copies of universe $B$.
These copies can be connected by replica wormholes, which should be included in the path integral. 
On a replica wormhole, the product of overlaps on the right hand side has an expression in terms of correlation functions of local operators via the state-operator correspondence \cite{WIP}.
By including the effect of such wormholes, we arrive at the  expression for the entropy $S(A)$
\be
    S(A)= \min
    \begin{cases}
    S_{\beta} (B) , \\
    \underset{C}{\text{min ext}} \left[ \phi[\partial \overline{C}]+  S_{\beta}[\overline{C}] -S_{\text{vac}}[\overline{C}] \right] ,
    \end{cases}
\label{eq:complement-islands}
\ee
where $S_\beta$ is the thermal state CFT entropy at inverse temperature $\beta$, $S_{\text{vac}}$ is the vacuum CFT entropy, and $\phi$ is the JT dilaton field.
Note that one of the important points raised in \cite{WIP} about this formula is the fact that it is the complement $\overline{C}$ of the island region $C$ which appears. In order to see the reason, It is useful to  write the right hand side, 
\be 
\phi[\partial \overline{C}]+  S_{\beta}[\overline{C}] -S_{\text{vac}}[\overline{C}]= \phi[\partial AC] +S_{\Psi}[AC] -S_{{\rm vac}} [AC] ,\label{eq:genen}
\ee
where $AC$ is the union of the Cauchy slice of the non gravitating universe $A$ and the island $C$ in the gravitating universe $B$, and $S_{\Psi}[AC]$ is the entropy of the pure state $| \Psi \ra$ defined in  \eqref{eq:HHstate} on $AC$. Thus, the right hand side of \eqref{eq:genen} can be identified with the generalized entropy which appears in island formula.  Also, we have in mind that  $C$ is some connected island region, but $\overline{C}$ could be disconnected depending on the topology of the spatial slice.  In de Sitter, we have a compact spatial slice so this point is not so crucial, but it will play more of a role in the universal cover of the black hole.   Also, as discussed in detail in \cite{WIP}, both candidate expressions appearing in \eqref{eq:complement-islands} are free of UV divergences.

\subsection{Two dimensional de Sitter JT gravity}

We now turn to the theory on universe $B$, which is 2d de Sitter JT gravity coupled to a quantum CFT.
The effective action is
\be 
-\log Z_B = \f{\phi_{0}}{16\pi G_N}\int \s{-g}R +\f{1}{16\pi G_N}\int \s{-g} \Phi \left( R -\f{2}{L^2}\right) - \log Z_{\text{CFT}} , 
\label{eq:dSJT}
\ee
where $\log Z_{\text{CFT}}$ is an effective action for quantum matter fields, $\Phi$ is the dilaton, and the total gravitational action is what we previously called $I_{\text{grav}}$.\footnote{We have not explicitly written the usual Gibbons-Hawking-York boundary term, but it is implicit in our expressions.}  This theory describes the near-horizon gravitational dynamics of an ``extremal"\footnote{By extremal, we do not mean that the black hole carries some maximal amount of charge or rotation such that adding more would produce a naked singularity.  Instead, we mean a static de Sitter black hole of near-maximal size, with horizon approaching the cosmological horizon of the spacetime.  This should be contrasted with the use of ``extremal" in the flat space and AdS contexts, where we do have in mind a charged or rotating black hole.} Schwarzchild-de Sitter black hole in higher dimensions. 
Beginning from the 4d Schwarzchild-de Sitter black hole \cite{Gibbons:1977mu},
\be
 ds^2 = -f(r) dt^{2} +\f{dr^{2}}{f(r)}+ r^2 (d\theta^{2}+ \sin^{2} \theta d\phi), \quad f(r)=1-\f{2M}{r} -\f{\Lambda r^{2}}{3} ,
\ee
notice that (for $M < 1/3\sqrt{\Lambda}$) there are two horizons corresponding to the two zeros of $f(r)$.
The smaller of these, $r_+$, is the black hole horizon.
The larger, $r_{++}$, is the cosmological horizon, associated with the fact that observers in de Sitter space can only observe a subset of the full spacetime.
If we send $M \to 1/3\sqrt{\Lambda}$, we will send $r_+ \to r_{++}$.
In this limit, the region between $r_+$ and $r_{++}$ becomes very small, and in a near-horizon limit becomes the Nariai spacetime dS$_2 \times S^2$ \cite{10026018884,40018660490}.
Dimensional reduction on the transverse sphere yields dS$_2$ JT gravity, in the same way that AdS$_2$ JT gravity arises from the dimensional reduction of a near-horizon limit for 4d flat space or AdS near-extremal black holes. The first term of the action \eqref{eq:dSJT} corresponds to the entropy of the extremal black hole, and the second term captures the deviations away from extremality. 
Two dimensional de Sitter JT gravity has been studied recently \cite{Maldacena:2019cbz,Cotler:2019nbi} as part of recent general developments concerning the information paradox in 2d gravity \cite{Saad:2019lba,Marolf:2020xie,Almheiri:2020cfm}.

The equation of motion for the metric is obtained by varying the action \eqref{eq:dSJT} with respect to the dilaton $\Phi$:
\be 
R-\f{2}{L^2}=0 .
\label{eq:metricEOM}
\ee
In two dimensions, this implies that the full Ricci tensor is proportional to the metric
\begin{equation}
    R_{ab} =\f{g_{ab}}{L^{2}} .
\end{equation}
Varying \eqref{eq:dSJT} with respect to the inverse metric, we find an equation of motion for the dilaton:
\be 
-\nabla_{a}\nabla_{b} \Phi+g_{ab} \nabla^{2} \Phi+  \f{g_{ab}}{L^{2}}\Phi =8\pi G_N \avg{T_{ab}} ,
\label{eq:dilatonEOM}
\ee
where $\avg{T_{ab}}$ is the expectation value of the stress energy tensor coming from the effective matter action $\log Z_{\text{CFT}}$.
The only difference between \eqref{eq:dilatonEOM} and the analogous expression in the AdS case is the sign of the last term on the left hand side.

\section{Semiclassical solution}\label{sec:semiclassical}
We work in Lorentzian signature and solve the equations of motion \eqref{eq:metricEOM} and \eqref{eq:dilatonEOM}.
The metric equation \eqref{eq:metricEOM} simply fixes the geometry to have constant positive curvature, so it is locally dS$_2$.
To find a semiclassical solution, all that remains is to solve the dilaton equation \eqref{eq:dilatonEOM} on de Sitter space. 
In what follows, we will fix $L=1$.
In conformal gauge and lightcone coordinates $x^\pm$, the metric is of the form 
\be 
ds^{2} = -e^{2\omega}dx^{+}dx^{-}, \quad x^{\pm} =\tau \pm \theta , \quad e^{-2\omega} = \cos^{2}\tau .
\ee 
We have defined global dS$_2$ coordinates $(\tau,\theta)$ with ranges $\tau \in (-\frac{\pi}{2},\frac{\pi}{2})$ and $\theta \in (-\frac{\pi}{2}, \frac{3\pi}{2})$.\footnote{The somewhat nonstandard range for $\theta$ is chosen to ensure our cosmological horizons do not lie on the boundary of the Penrose diagram, where the left and right edges are identified since the spatial section is topologically a circle.}

In lightcone coordinates, \eqref{eq:dilatonEOM} splits into three component equations.
The $++$ and $--$ components are given by
\be
e^{2\omega} \p_{+} \left[ e^{-2\omega}\p_{+} \Phi \right] =-8\pi G_N \la T_{++}\ra, \quad e^{2\omega} \p_{-} \left[ e^{-2\omega}\p_{-} \Phi \right] =-8\pi G_N \la T_{--} \ra, \label{eq:eq12}
\ee
and the $+-$ component is
\be 
-e^{2\omega}\Phi +2 \p_{+}\p_{-} \Phi =16\pi G_N \la T_{+-} \ra . \label{eq:eq3}
\ee
The first two equations \eqref{eq:eq12} are independent of the signature of cosmological constant;
only the third equation \eqref{eq:eq3} is modified from that of AdS JT gravity, where the first term on the left hand side has opposite sign.

\subsection{The sourceless solution}
\begin{figure}[t]
    \centering
    \includegraphics[scale=.3]{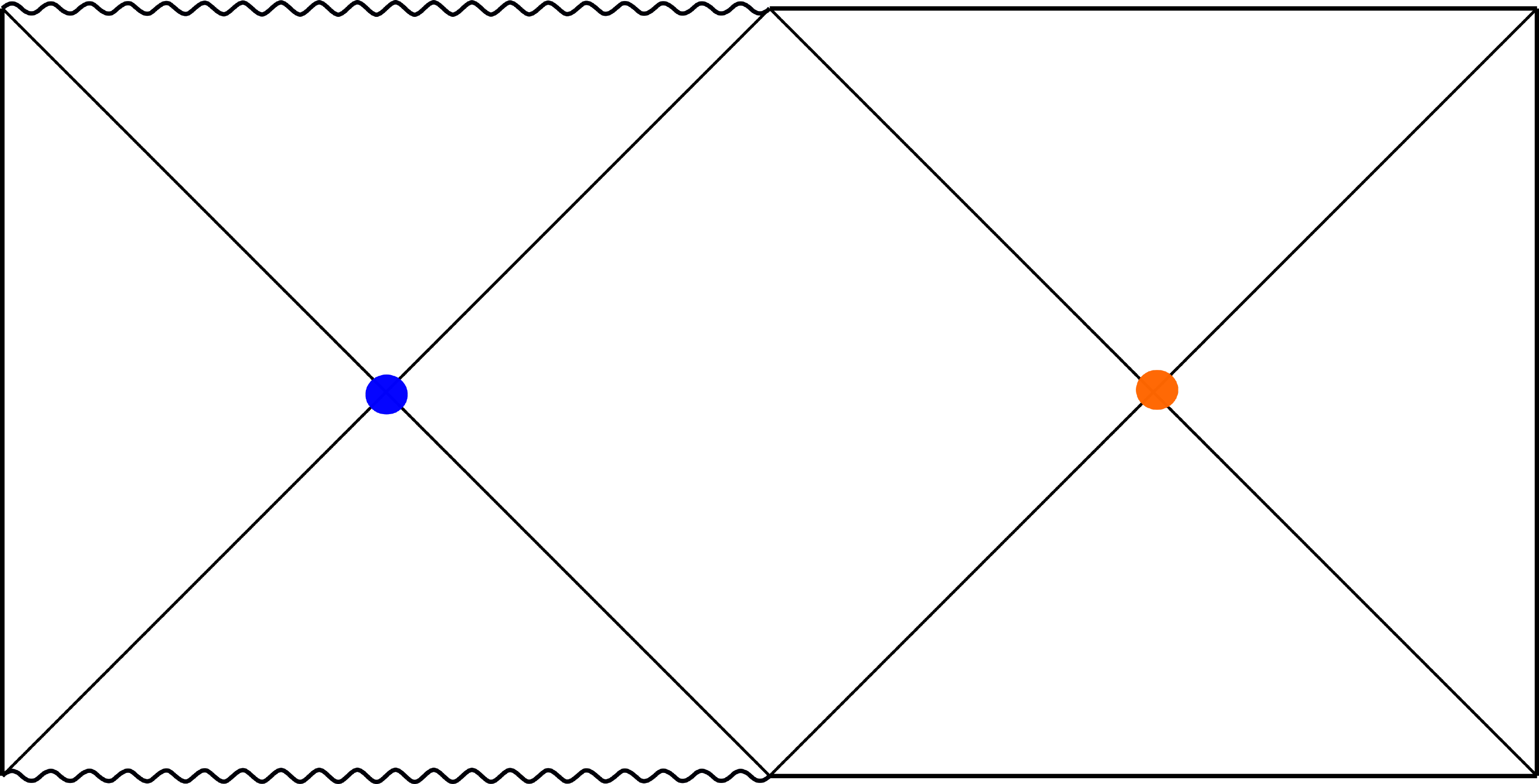}
    \caption{\small{The Penrose diagram of the black hole with the dilaton profile \eqref{eq:generalsourceless} ($\zeta=0$). The blue dot is the event horizon of the black hole at $\theta=\pi$, and the orange dot is the cosmological horizon at $\theta=0$.
    In this work, we will use a somewhat nonstandard convention where $\theta$ increases from right to left.
    }}
    \label{fig:2dschbh}
\end{figure}

Now let us solve the equation \eqref{eq:dilatonEOM} when $\la T_{++}\ra =\la T_{--} \ra=\la T_{+-} \ra=0$. 
There are two independent solutions to this second order linear differential equation, and the general linear combination is
\be 
 \Phi_0(\tau,\theta) = \zeta \tan \tau + \alpha \f{\cos \theta}{\cos \tau}, 
\label{eq:generalsourceless}
\ee
for some constants $\zeta$ and $\alpha \geq 0$.
We will be interested in the solution where $\zeta = 0$.
In this case, there are two horizons at $(\tau, \theta)=(0,0)$ and $(\tau, \theta)=(0,\pi)$, corresponding to the points where the derivatives of $\Phi_0$ vanish. 
The first of these points corresponds to the cosmological horizon, and the second corresponds to the black hole horizon.  
Indeed, their entropies are given by the dilaton values 
\be
S_{CH}=\phi_{0}+\Phi_0(0,0), \quad S_{BH}=  \phi_{0}+\Phi_0(0,\pi) ,
\ee
and they satisfy $\Phi_0(0,0)>\Phi_0(0,\pi)$. The Penrose diagram of this spacetime is shown in Fig.~\ref{fig:2dschbh}. 
The appearance of these two horizons can be naturally understood from the point of view of dimensional reduction from the four dimensional Schwarzschild-de Sitter black hole in the Nariai limit.

\subsection{The solution with source}

Now let us solve the full equation of motion \eqref{eq:dilatonEOM}, including the stress energy source term.
A similar geometry in AdS was discussed in \cite{Bak:2018txn}. Here we generalize their result to the de Sitter case.

In a curved background, the stress tensor expectation values receive corrections from the Weyl anomaly (which contributes to $\avg{T_{\pm\pm}}$) and, after requiring stress tensor conservation (which relates $\avg{T_{+-}}$ to $\avg{T_{\pm\pm}}$), we have
\be
\la T_{\pm \pm} \ra = \f{c}{12\pi} \left(\p_{\pm}^{2} \omega -(\p_{\pm}\omega)^{2} \right) +\tau_{\pm \pm},  \qquad \la T_{+-}\ra = -\f{c}{12\pi}\p_{+}\p_{-}\omega,
\ee
where $\tau_{\pm \pm}$ is the expectation value in flat space.
We are interested in a thermal state, so $\tau_{\pm\pm}$ is the expectation value in a thermal state on flat space, which is related by the exponential map to the vacuum state on flat space.
So, there is a contribution from the Schwartzian derivative and a contribution from the Casimir energy since our CFT is on a circle rather than a line, summing to
\begin{equation}
    \tau_{\pm\pm} = \frac{c}{24} \left( \frac{2\pi}{\beta} \right)^2 - \frac{c}{48\pi} .
\end{equation}
The Casimir energy cancels the contribution from the Weyl anomaly, yielding
\be
\la T_{\pm \pm }\ra = \f{c}{24\pi} \left(\f{2\pi} \beta \right)^{2},\quad  \la T_{+-}\ra =-\f{c}{48\pi \cos^{2} \tau}.
\ee
Then the total solution is 
\be 
\Phi (\tau, \theta)= \alpha\f{\cos \theta}{\cos \tau} - \f{K}{2} (\tau \tan \tau+1)+\f{cG}{3}, \quad K \equiv \f{cG}{3} \left(\f{2\pi}{\beta}\right)^{2}. \label{eq:totaldil}
\ee
Though we have recorded the de Sitter solution where we imagine the Cauchy slice is a compact circle, we can also adapt this solution to the case where we imagine passing to the universal cover of a de Sitter black hole spacetime.
In the universal cover, the Cauchy slice becomes noncompact, and the CFT stress tensor from flat space will no longer have a Casimir energy contribution $-\frac{c}{48\pi}$.
The solution with source in this universal cover situation is then related to the compact situation by the simple replacement $K \to K'$, where
\begin{equation}
    K' \equiv 4\pi K + \frac{2cG_N}{3} . \label{eq:modified}
\end{equation}

\subsection{Imposing asymptotic de Sitter boundary conditions}

Thus far, the coefficient $\alpha$ in the sourceless part of the solution \eqref{eq:totaldil} has been arbitrary.
However, enforcing asymptotically de Sitter boundary conditions will determine $\alpha$ in terms of the CFT temperature $\beta$ and the entropy of the cosmological horizon (equivalently, the dimension of the de Sitter quantum gravity Hilbert space).
We will enforce these boundary conditions by demanding at late time $\tau \rightarrow \f{\pi}{2}$, the dilaton profile is asymptotically equivalent to that of pure de Sitter.
In order to do so, we start from the following sourceless dilaton profile,
\be
\Phi_{0} (\tau, \theta)= \f{\bar{\phi}}{2}\left[\left(b+\f{1}{b}\right)\f{\cos \theta}{\cos \tau} -\left(b-\f{1}{b} \right)\tan \tau \right] ,
\label{eq:sourceless}
\ee
which is just a re-expression of \eqref{eq:generalsourceless}.
Our motivation for parametrizing the pure de Sitter dilaton profile in this manner comes from the form of the Milne coordinate patch, which covers the future lightcone of the cosmological horizon at $(\tau,\theta) = (\tau_0,0)$, defined by the points where $\partial_\theta \Phi_0 = \partial_\tau \Phi_0 = 0$ in \eqref{eq:sourceless}:
\be 
\sin \tau_{0} =\f{b^{2}-1}{b^{2}+1} .
\label{eq:CHsourceless}
\ee
We would like to match the late time dilaton in this region which covers spacelike future infinity.
To see this (we leave the details to an appendix), notice that the relations\footnote{The set of coordinate transformations \eqref{eq:coordtrans} is obtained by first embedding de Sitter space in global coordinates  into a three dimensional hyperboloid, applying an $SL(2,\mathbb{R})$ transformation, and then pulling back by the Milne coordinates.}
\be
r=\f{1}{2}\left(b+\f{1}{b}\right)\f{\cos \theta}{\cos \tau} -\f{1}{2} \left(b-\f{1}{b} \right)\tan \tau, \quad \tanh t= \f{1}{2}\left(b+\f{1}{b}\right)\f{\sin \tau}{\sin \theta}-\f{1}{2}\left(b-\f{1}{b}\right)\f{1}{\tan\theta} ,
\label{eq:coordtrans}
\ee
one can identify the future light cone of the cosmological horizon \eqref{eq:CHsourceless} with the aforementioned Milne patch of de Sitter space,
which has the metric and dilaton profile 
\be 
ds^{2} =-(1-r^{2}) dt^{2} +\f{dr^{2}}{1-r^{2}}, \quad \Phi= \bar{\phi}r, \quad r \in [1, \infty) .
\label{eq:static}
\ee
In these coordinates, the manifest cosmological horizon at $r=1$ matches the value of our sourceless dilaton at $(\tau_0,0)$ as expected.
Motivated by this, we fix the coefficient $\alpha$ in the total solution (\ref{eq:totaldil}) by expanding \eqref{eq:sourceless} and \eqref{eq:totaldil} around $\tau = \frac{\pi}{2}$ and matching the leading divergent terms.
This leads to expressions for $K$ and $\alpha$ in terms of $\bar{\phi}$ and $b$:
\be 
\f{\pi K}{4}=\f{\bar{\phi}L}{2}\left(b-\f{1}{b}\right),\quad \alpha=\f{\bar{\phi}L}{2}\left(b+\f{1}{b}\right).
\label{eq:defofb}
\ee
These relations ensure that at late time $\tau \rightarrow \f{\pi}{2}$ the total solution \eqref{eq:totaldil} can be approximated by the sourceless, pure de Sitter solution \eqref{eq:sourceless}.

\section{Penrose diagram}\label{sec:penrose}

Before turning to entropies, we pause to study the causal structure of the backreacted spacetime (Fig.~\ref{fig:causalsh}).
In particular, we specify the locations of the singularity, spacelike infinity, and horizons, all of which can be extracted from the dilaton profile.
Here we reproduce the total dilaton profile\footnote{We have absorbed $cG_N/3$ in the last term of (\ref{eq:totaldil}) into the definition of $\phi_{0}$. } 
\be 
\phi (\tau, \theta)= \phi_{0} + \Phi(\tau, 
\theta)= \phi_{0}+\f{\bar{\phi}L}{2}\left[\left(b+\f{1}{b}\right) \f{\cos \theta}{\cos \tau}  -\f{2}{\pi}\left(b-\f{1}{b}\right) (\tau \tan \tau+1) \right], \quad  \label{eq:totaldil2}
\ee
where we have included the constant $\phi_0$ which appears in the ground state entropy term in JT gravity,\footnote{We note that it is this complete dilaton $\phi$ which enters the calculation of the entropy, and it is zeroes of this dilaton that signal singularities in the 2d spacetime.} and we have also made the substitutions \eqref{eq:defofb} in \eqref{eq:totaldil}. In particular the parameter $b$ depends on the entanglement temperature $\beta$.

\subsection{Singularity and spacelike infinity} 

In the dimensional reduction from 4d, the dilaton emerges as a measure of the radius of the transverse sphere.
A singularity therefore corresponds to a region where the dilaton \eqref{eq:totaldil2} is vanishing, or equivalently where \eqref{eq:totaldil} is becoming very negative, since we imagine $\phi_0$ is a large constant.
Since $K$ can be expressed in terms of the central charge and Newton's constant \eqref{eq:totaldil}, and $\alpha$ should be positive to ensure positivity of \eqref{eq:sourceless}, the combinations $b\pm 1/b$ are always non-negative.
This means that the condition for $\phi$ to vanish at some point as $\tau \to \frac{\pi}{2}$ is
\begin{equation}
     \cos\theta \leq 1-\frac{2}{b^2+1} ,
\label{eq:singularity-range}
\end{equation}
where we have simply compared the divergent pieces of the first and second terms in the square brackets of \eqref{eq:totaldil2}.
Since both $b \pm 1/b$ are non-negative, we must have $b \geq 1$, which implies that $\frac{\pi}{2} < \theta < \frac{3\pi}{2}$ is the smallest range for which we encounter a singularity as $\tau \to \frac{\pi}{2}$ (Fig.~\ref{fig:causalsh}).
As $b$ increases (which is achieved by increasing the CFT temperature $1/\beta$), the range of $\theta$ for which we encounter a singularity as $\tau \to \frac{\pi}{2}$ also increases.
Of course, for a fixed value of $\theta$ which is in the complement of \eqref{eq:singularity-range}, we instead reach future spacelike infinity by taking $\tau \to \frac{\pi}{2}$.
As the dilaton \eqref{eq:totaldil2} is symmetric under $\tau \to -\tau$, our results here also apply to the past singularity and past spacelike infinity, but with $\tau \to -\frac{\pi}{2}$.

\subsection{Black hole apparent and event horizons}

The event horizons of the black hole travel along null lines from the intersections of the singularity with the boundary of the Penrose diagram, when the inequality \eqref{eq:singularity-range} is saturated.
Notice that since the left event horizon and the right event horizon do not intersect for any $b > 1$, there is a causal shadow region between them which forms due to the backreaction as soon as $\beta < \infty$.

\begin{figure}[t]
    \centering
    \includegraphics[scale=.3]{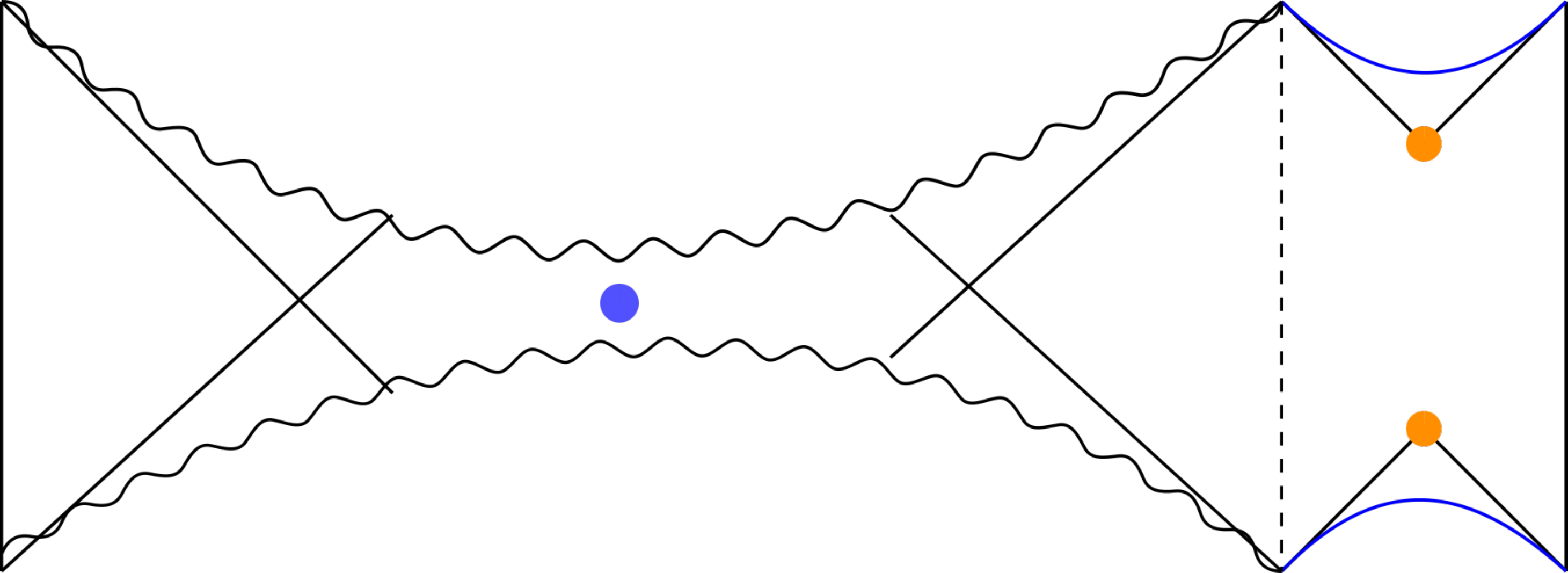}
    \caption{\small{The Penrose diagram of the backreacted black hole. 
    As we increase the 
    CFT temperature, the black hole interior region gets longer.  The blue dot represents the apparent horizon of the black hole, which differs from the event horizons. The orange dots are the cosmological apparent horizons, which are shown as overlapping the cosmological event horizons, but in general there is a slight difference between their positions.  In the high temperature limit, they coincide at the diagram boundary.}}
    \label{fig:causalsh}
\end{figure}

In order to see this, first observe that the singularity meets the future boundary  $\tau=\f{\pi}{2}$ at $\theta_{L} = 2\pi -\theta_{0}$ and $\theta_{R} = \theta_{0}$, where $\theta_{0} $ saturates the bound \eqref{eq:singularity-range}, and $   0\leq\theta_{0}  \leq \f{\pi}{2} $.  The left future event horizon $\mathcal{H}_{L}$  of the black hole is the null line  starting from $( \tau, \theta) =(\f{\pi}{2}, \theta_{L})$, and  similarly  the right future event horizon is from $( \tau, \theta) =(\f{\pi}{2}, \theta_{R})$
\be 
\mathcal{H}_{L}:  \theta= \theta_{L} + \left(\tau-\f{\pi}{2} \right), \quad  \mathcal{H}_{R}: \theta= \theta_{R} -\left(\tau-\f{\pi}{2} \right) . \label{eq:horizons}
\ee
There are similar expression for past event horizons (see Fig.~\ref{fig:causalsh}).
The future and past event horizons meet at $\tau=0$ slice. 
From \eqref{eq:horizons}, we  find that the intersections (the bifurcation surfaces) are located at $\theta_\pm$ which satisfy
\begin{equation}
    \theta_{+} = \f{3\pi}{2}-\theta_{0}, \quad \theta_{-}= \f{\pi}{2}+ \theta_{0} .
\label{eq:BH-event-horizon}
\end{equation}
The dilaton takes equal values at these two points,
\begin{equation}
    \phi(0,\theta_\pm) = \phi_0 - \bar{\phi} L \left[1 + \frac{2}{\pi } \left(b-\f{1}{b} \right) \right] .
\end{equation}
In the high temperature limit $\beta \rightarrow 0$ (equivalently, $b \to \infty$), the intersections of the singularity with the diagram boundary are moving into the corners of the diagram $\theta_{0} \rightarrow 0$, consistent with the range \eqref{eq:singularity-range} encompassing the entire coordinate range of $\theta$ (Fig.~\ref{fig:causalsh}).
In this limit, the event horizon  asymptotes to the null lines at the corners of the Penrose diagram.

However, we cannot actually reach the limit
$\theta_{0} =0$
 while maintaining semiclassical control over the solution.
To see this, consider the location in $\theta$ where the singularity is closest to the $\tau = 0$ surface.
In other words, let $\tau(\theta)$ be the location of the singularity, i.e. the curve satisfying  $\phi (\tau(\theta), \theta)=0, \tau(\theta)> 0$. Then consider the minimum of this function $\tau(\theta)$.
We deduce that the singularity is closest to the $\tau = 0$ (or $\tau(\theta)$ takes the minimum value) at $\theta = \pi$.  This is because  the singularity is symmetric about $\theta = \pi$, and  $\tau (\theta)$ is monotonically decreasing in the window $ \theta_{R} <\theta <\pi$.
As we increase temperature, the closest singular point comes down toward $(\tau,\theta) = (0,\pi)$.
In fact, there is a critical value of $b$ (equivalently, $1/\beta$) where the dilaton \eqref{eq:totaldil2} at this point is zero:
\begin{equation}
    b_{\text{crit}} = \frac{\sqrt{\pi^2 \phi_0^2-\left(\pi^2-4\right) L^2 \bar{\phi}^2}+\pi  \phi_0}{(2+\pi )
   L \bar{\phi}} .
\end{equation}
Above this value of $b$, the future and past singularities are joined, and the black hole horizons are absorbed into the singularity. This process has been studied by Bousso \cite{Bousso:1998bn}.
Therefore, it is hard to calculate entropies beyond this point, since the semiclassical description is breaking down. 
A full treatment of the entropy beyond this regime requires quantum JT gravity.
Fortunately, we will be able to see all effects we are interested in long before reaching this point.
This is because $b_{\text{crit}}$ increases linearly with $\phi_0$, and we are free to make the extremal entropy as large as we like (up to the bound on entropy determined by cosmological constant via the dimensional reduction to JT gravity of the higher dimensional near-extremal black hole).  Thus, when we speak about ``high temperature" limits, it should be understood in the sense of being much larger than 1 but still smaller than $\phi_0$.
We will return to this point in Sec.~\ref{sec:island}.

We have specified the location of  the event horizon of the black hole from the behavior of the singularity, and also understood its asymptotics for $\beta \to 0$. However, the event horizon does not extremize the dilaton (or, equivalently, the area of co-dimension 2 surfaces in the dimensionally lifted picture).  Black hole entropy should be associated to loci that extremize the dilaton.
Such loci correspond to apparent horizons, which are generally located inside the event horizon, although for a stationary black hole, the apparent horizon coincides with the event horizon.
Extrema of the dilaton are governed by the equations $\partial_\theta \phi = \partial_\tau \phi = 0$.
The $\theta$ equation implies $\theta = 0$ or $\theta = \pi$, and only $\theta = \pi$ lies between the event horizons.
Given $\theta = \pi$, the $\tau$ equation is
\be 
\left(b+\f{1}{b}\right) \sin \tau +\left(b-\f{1}{b}\right) (\tau +\sin \tau \cos \tau)=0,
\ee
so we have found the apparent horizon at $(\tau,\theta)=(0,\pi)$.
It is easy to verify that this is the unique solution to the above equation for any $b \geq 1$ by noticing that the $\tau$ derivative of the left hand side is always positive on $\tau \in [-\frac{\pi}{2},\frac{\pi}{2}]$, so the function itself is monotone increasing on this interval, which implies there is only one zero.
The dilaton value here is\footnote{It may seem strange to define the black hole entropy as the dilaton value at the apparent horizon rather than at the event horizon.  We do so because it is this dilaton value which will enter in our computation of the entropy via the island formula in Sec.~\ref{sec:island}.  There are also situations where the apparent horizon is in fact the correct measure of coarse-grained black hole entropy \cite{Engelhardt:2017aux}.}
\be
S_{BH} = \phi(0,\pi)=\phi_{0}- \f{\bar{\phi}L}{2} \left[\left(b+\f{1}{b}\right)- \f{2}{\pi}\left(b-\f{1}{b}\right)\right] \label{eq:BHarea}
\ee
In the high temperature limit $\beta \rightarrow 0$, recalling $b \sim \f{1}{\beta^{2}}$, the black hole entropy is decreasing as we increase the temperature. 
We interpret this as the evaporation of the black hole.
We will see in Sec.~\ref{sec:island} that the generalized entropy is dominated by this horizon area, which results in reproduction of the Page curve. 

\subsection{Cosmological apparent and event horizons}

Our spacetime also supports cosmological apparent horizons, also defined by extremization of the dilaton.
Like the black hole apparent horizon and the black hole event horizons, the cosmological apparent horizons need not coincide with the cosmological event horizons, and in general lie within the causal diamond whose top and bottom corners are  the cosmological event horizons.

To locate the cosmological apparent horizons, we must write the $\tau$ extremization equation with the choice $\theta = 0$: 
\be 
\left(b+\f{1}{b}\right) \sin \tau -\f{2}{\pi}\left(b-\f{1}{b}\right) (\tau +\sin \tau \cos \tau)=0 .
\label{eq:deftau0}
\ee
There are three solutions of \eqref{eq:deftau0} in general, which are located at $\tau = 0$ and $\tau = \pm \tau_0$.
The appearance of the nonzero solutions  $\tau=\pm \tau_{0}$ is not immediate with $b>1$.
There is a finite window of $b > 1$ where the only solution of \eqref{eq:deftau0} is $\tau=0$.
To understand this, we expand \eqref{eq:deftau0} around $\tau = 0$ and obtain a cubic equation.
There is a triple root at zero when
\begin{equation}
    b_{\text{triple}} = \sqrt{\frac{4+\pi}{4-\pi}} ,
\label{eq:triple-root-b}
\end{equation}
and this represents the point (as a function of $b$) where two imaginary roots of the cubic equation become real.
So, between $1 \leq b \leq b_{\text{triple}}$, the only solution of \eqref{eq:deftau0} is $\tau = 0$.
For $b > b_{\text{triple}}$, there are three solutions, and the positive one defines $\tau_0$.
Unfortunately, the transcendental equation \eqref{eq:deftau0} is hard to solve analytically, so we can only evaluate the dilaton numerically at the cosmological apparent horizons.

Of course, due to the growth of the singularity \eqref{eq:singularity-range} at $b>1$, the cosmological event horizons at $(\pm \tau_{e}, 0)$ must move away from $(\tau,\theta) = (0,0)$ immediately as $b>1$.
Again by drawing a null line $\tau= \f{\pi}{2} +(\theta-\theta_{0})$ which starts from one of the endpoints of the black hole singularity $(\tau, \theta) =(\f{\pi}{2}, \theta_{R})$, we get
\begin{equation}
    \tau_{e} = \frac{\pi}{2}-\theta_{0} .
\end{equation}
Since the dilaton \eqref{eq:totaldil2} is symmetric under $\tau \to -\tau$, it takes equal values at these horizons.
\begin{equation}
    \phi(\pm\tau_{e},0) = \phi_0 + \frac{\bar{\phi}L}{2\pi b^2} \left[ (b^2-1)^2 \arccos \left( 1 -\frac{2}{b^2+1} \right) - 2b^3 + 2\pi b^2 + 2b \right] .
\end{equation}
Though we have imposed asymptotically de Sitter boundary conditions, the value of the dilaton at both the cosmological apparent\footnote{This can be checked by solving the transcendental equation \eqref{eq:deftau0} numerically and then evaluating \eqref{eq:totaldil2} at $(\tau_0,0)$.} and event horizons changes as we tune $\beta$.
But we can say is that the entropy of the cosmological event horizon has the upper bound
\begin{equation}
    \phi(\pm \tau_{e},0)|_{b=1} = \phi(\pm \tau_{e},0)|_{b\to\infty} = \phi_0 + \bar{\phi} L ,
\end{equation}
which is reached at both the zero temperature and infinite temperature limits.

\subsection{Universal covering space}

\begin{figure}[t]
    \centering
    \includegraphics[scale=.5]{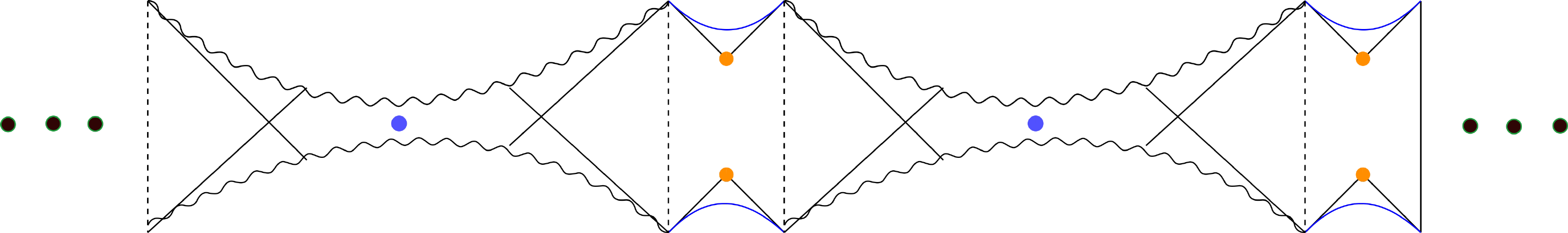}
    \caption{\small{The universal covering space of the Penrose diagram of the backreacted black hole. Black dots indicate that we can continue this pattern indefinitely.}}
    \label{fig:maximal}
\end{figure}

We want to allow for the possibility of passing to the universal cover of our black hole geometry (Fig.~\ref{fig:maximal}).
The metric itself trivially allows for such an extension, as the pure de Sitter metric we are working with has the Killing vector $\partial_\theta$, and thus we may decompactify the angular direction.\footnote{In higher dimensions, this procedure is more subtle and involves complicated coordinate transformations which keep the metric regular around bifurcation surfaces.  Due to the JT equations of motion, our metric is fixed and already has no coordinate singularities, so such issues do not arise.}
However, we must consider the behavior of the dilaton under this operation.
Fortunately, the dilaton \eqref{eq:totaldil2} depends on $\theta$ only through $\cos \theta$, and this implies the extended Penrose diagram (Fig.~\ref{fig:maximal}) is just an infinite sequence of the compact Penrose diagram (Fig.~\ref{fig:causalsh}), just as in the higher dimensional Schwarzschild-de Sitter black hole.
The statements we have made here are true for any value of the CFT temperature $1/\beta$, that is to say, for the full backreacted Penrose diagram.

\subsection{Summary}

As we increase the entanglement temperature, the structure of the spacetime gradually changes.  The interior of the black hole grows, and (if we could retain semiclassical control all the way to $b \to \infty$) reaches its maximal size when it takes up a coordinate range $\frac{\pi}{2} < \theta < \frac{3\pi}{2}$, which is half of the of total time slice, which has a coordinate range of $2\pi$.   During this process, the black hole apparent horizon remains at $\theta=\pi$, in the black hole interior.  The dilaton value at the apparent horizon decreases, which can be interpreted as ``evaporation'' of the black hole through entanglement with the auxiliary universe.

The locations of the cosmological apparent and event horizons, on the other hand, change with $\beta$.  The apparent horizon $(\tau, \theta)=(\tau_{0},0)$ goes to future infinity $\tau_{0} \rightarrow \f{\pi}{2}$ as $\beta \rightarrow 0$, and similarly for the future cosmological event horizon $\tau_{e}$. At first sight, the future light cone of the horizons appear to be shrinking in the Penrose diagram (Fig.~\ref{fig:causalsh}), since they are approaching the boundaries. 
However, the actual size of the horizon  measured by the dilaton is never decreasing.

\subsection{Comparison with AdS}

Previously, we studied the backreaction on an asymptotically anti-de Sitter black hole in the same setup as the one we consider in this paper \cite{WIP}.  One key difference here is that the de Sitter black hole entropy decreases in the $\beta \rightarrow 0$ limit instead of becoming constant as in  AdS.  In essence this is because de Sitter black holes can evaporate, while anti-de Sitter black holes come into equilibrium with their radiation. Another key difference in the de Sitter case will involve the nature of the entanglement island, which we study in Sec.~\ref{sec:island}. In the AdS case, the island almost 
coincided with the entire black hole interior which approached the AdS boundary and the entropy remained constant as we increased the temperature.  In the de Sitter case, we will show that the island is instead the complement to a very tiny region near the apparent horizon in the black hole interior.  An island  similar in spirit to the AdS situation appears if we instead consider the universal covering space of the de Sitter black hole.

\section{Islands in de Sitter}\label{sec:island}

Having specified the backreacted dilaton \eqref{eq:totaldil2}, and with an understanding of the Penrose diagram as a function of $\beta$ from Sec.~\ref{sec:penrose}, we now calculate the entanglement entropy $S_{A}$ of the state \eqref{eq:HHstate}. 
Schematically, this is given by 
\be
S_{A} = \min \left\{S_{\text{no-island}}, S_{\text{island}} \right\}
\ee
In the replica derivation of this formula, $S_{\text{no-island}}$ comes from the contribution of the fully disconnected saddle in the gravitational path integral.  This is given by the thermal entropy $S_{\beta}(B)$ of the CFT living on the gravitating universe B, so we have $S_{\text{no-island}} =S_{\beta}(B)$. 
On the other hand, $S_{\text{island}}$ comes from the  contribution of the fully connected replica wormhole, which is given by the minimum of the generalized entropy $S_{\text{gen}}[\overline{C}]$ for a spacelike interval $\overline{C}$ on the gravitating universe \eqref{eq:complement-islands}: 
\begin{equation}
S_{\text{island}}=\underset{\overline{C}}{\text{min ext}} \left[\phi[\partial \overline{C}] +S_{\beta} [\overline{C}] -S_{{\rm vac}} [\overline{C}] \right] .
\label{eq:gen-entropy-island-phase}
\end{equation}
(Throughout this section, we set $4G_N = 1$.) As stressed below \eqref{eq:complement-islands},  $\overline{C}$ is the complement of the island $C$ in a Cauchy slice of  the gravitating universe $B$. 
Let $u_{1}, u_{2}$ be two end points of the interval $\overline{C}$.  Here we employed an slightly generalized notation in order to include the cases where two endpoints may not lie in a fixed $\tau$ slice, i.e. if we denote $u_{1}:(\tau_{1}, \theta_{1})$ and $u_{2}:(\tau_{2}, \theta_{2}),$ in general we may have $\tau_{1} \neq \tau_{2}$. Then $\phi[\partial \overline{C}]$ in the above equation is the sum of the dilaton values at the two endpoints of the interval $\overline{C}$, 
\be 
\phi[\partial \overline{C}]= \phi(\tau_{1}, \theta_{1}) +\phi(\tau_{2}, \theta_{2}) .
\ee
The remaining terms $S_{\beta} [\overline{C}]$ and $ S_{{\rm vac}}[\overline{C}]$ represent the thermal field theory subregion entropy of a 2d CFT on a circle at inverse temperatures $\beta$ and $\infty$, respectively, and are given by (at large central charge $c$)\footnote{We use the holographic expressions \cite{Hubeny:2007xt} for thermal subregion entropies, though we expect all of our results to be insensitive to the particular CFT we pick.  As long as these functions are increasing with temperature at a reasonable rate, our results should be universal.  This is similar to the situation in \cite{Almheiri:2019qdq}, where a free fermion was used as a model.}
\begin{align}
S_{\beta} [\overline{C}] & =\f{c}{6} \log \left[\f{\beta}{\pi \varepsilon} \sinh \f{\pi}{\beta}(\theta_{2}-\theta_{1} +\tau_{2}-\tau_{1}) \right] +\f{c}{6} \log \left[\f{\beta}{\pi  \varepsilon} \sinh \f{\pi}{\beta}(\theta_{2}-\theta_{1} -(\tau_{2}-\tau_{1})) \right] , \label{eq:thermal-cft-entropy-subregion} \\
S_{\text{vac}} [\overline{C}] & =\f{c}{6} \log \left[\f{2}{\varepsilon} \sin \f{(\theta_{2}-\theta_{1} +\tau_{2}-\tau_{1})}{2} \right] +\f{c}{6} \log \left[\f{2}{\varepsilon} \sin \f{(\theta_{2}-\theta_{1} -(\tau_{2}-\tau_{1})) }{2}\right] \label{eq:vacuum-cft-entropy-subregion}.
\end{align}
where $\varepsilon$ is the ultraviolet cutoff. 

The intuition to keep in mind for these calculations is that, at high temperatures (where we expect to be in the Page phase of the entropy calculation), we search for a complement island $\overline{C}$ which is as small as possible, in order to have the island (and thus the entanglement wedge of universe $A$) be as large as possible.
We will see several examples of this in the subsections which follow; the complement islands at high temperature will either be small or bounded in size.

\subsection{Islands in the sourceless solution}\label{subsection:sourcelessisland}

It is instructive to first specify the locations of entanglement islands in the geometry without backreaction, where $\phi= \phi_{0}+\Phi_{0} (\tau, \theta)$.  Technically speaking, this calculation is invalid by assumption, since we are including the thermal field theory contributions to the generalized entropy while  we are neglecting the effect of the backreaction of the state $|\Psi \ra$  defined in \eqref{eq:HHstate} on the dilaton $\phi$ through its stress tensor expectation value $ \la \Psi |T_{\mu\nu}|\Psi \ra$.  Instead, we are 
using a sourceless dilaton profile $\Phi_{0}$ which assumes zero stress tensor contribution in the gravitational equations of motion.
However, let us proceed anyway, in order to understand the sorts of solutions to the problem of extremizing the generalized entropy \eqref{eq:gen-entropy-island-phase} which we may encounter in the more  complete setting.

In this case, we can assume that the island $C$ is on the $\tau=0$ Cauchy slice, as all dilaton extrema are located on that slice.
The generalized entropy is reduced to a function of two parameters,  $\theta_{1}$ and $\theta_{2}$ (with $\theta_2>\theta_1$), which specify the endpoints of the island. Therefore, for the sourceless solution, the generalized entropy function which appears in the island formula \eqref{eq:gen-entropy-island-phase} is\footnote{We set $L=1$ in this subsection.}
\be 
 S_{{\rm gen}}(\theta_{1}, \theta_{2}) = 2\phi_{0} + \bar{\phi} (\cos \theta_{1} +\cos \theta_{2}) + \f{c}{3} \log \left[ \f{\beta}{\pi} \sinh \f{\pi}{\beta}  (\theta_{2}-\theta_{1})\right]-\f{c}{3} \log \left[2\sin\f{(\theta_{2}-\theta_{1})}{2}\right].
\ee
Notice the cancellation of the ultraviolet cutoff $\varepsilon$ between the two terms $S_{\beta} [\overline{C}]$ and $S_{\text{vac}} [\overline{C}]$. This occurs because universe $A$ is disjoint from universe $B$.

All possible islands  can be found by solving $\partial_{\theta_{1}} S_{{\rm gen}} =\partial_{\theta_{2}} S_{{\rm gen}}=0$, which  requires 
\begin{equation}
\sin \theta_{1} +\sin \theta_{2}=0,
\label{eq:sourceless-island-condition-1}
\end{equation}
\begin{equation}
    \frac{c}{3} \cot \frac{\theta_2-\theta_1}{2} - \frac{2\pi c}{3\beta} \coth \frac{\pi}{\beta} (\theta_2-\theta_1) - \bar{\phi} (\sin\theta_1 - \sin\theta_2) = 0 .
\label{eq:sourceless-island-condition-2}
\end{equation}
We will find three types of solutions to these equations, which we will call Type I, II, and III.   Again instead of specifying these island themselves, it is convenient to specifying their complements $\bar{C}$ on the Cauchy slice, since we evaluate the generalized entropy on these complements (see Fig.~\ref{fig:nonIsland}).  We will see that the complements of  Type I islands are localized near the cosmological horizon, the complements of  Type II islands run between the cosmological horizon and the black hole horizon, and the complements of
type III islands are localized near the black hole horizon. 
These three classes correspond to the solutions of \eqref{eq:sourceless-island-condition-1}, and within each type we fix the precise island by solving \eqref{eq:sourceless-island-condition-2}.

\subsubsection*{Type I islands}

\begin{figure}
\centering
\includegraphics[width=6cm]{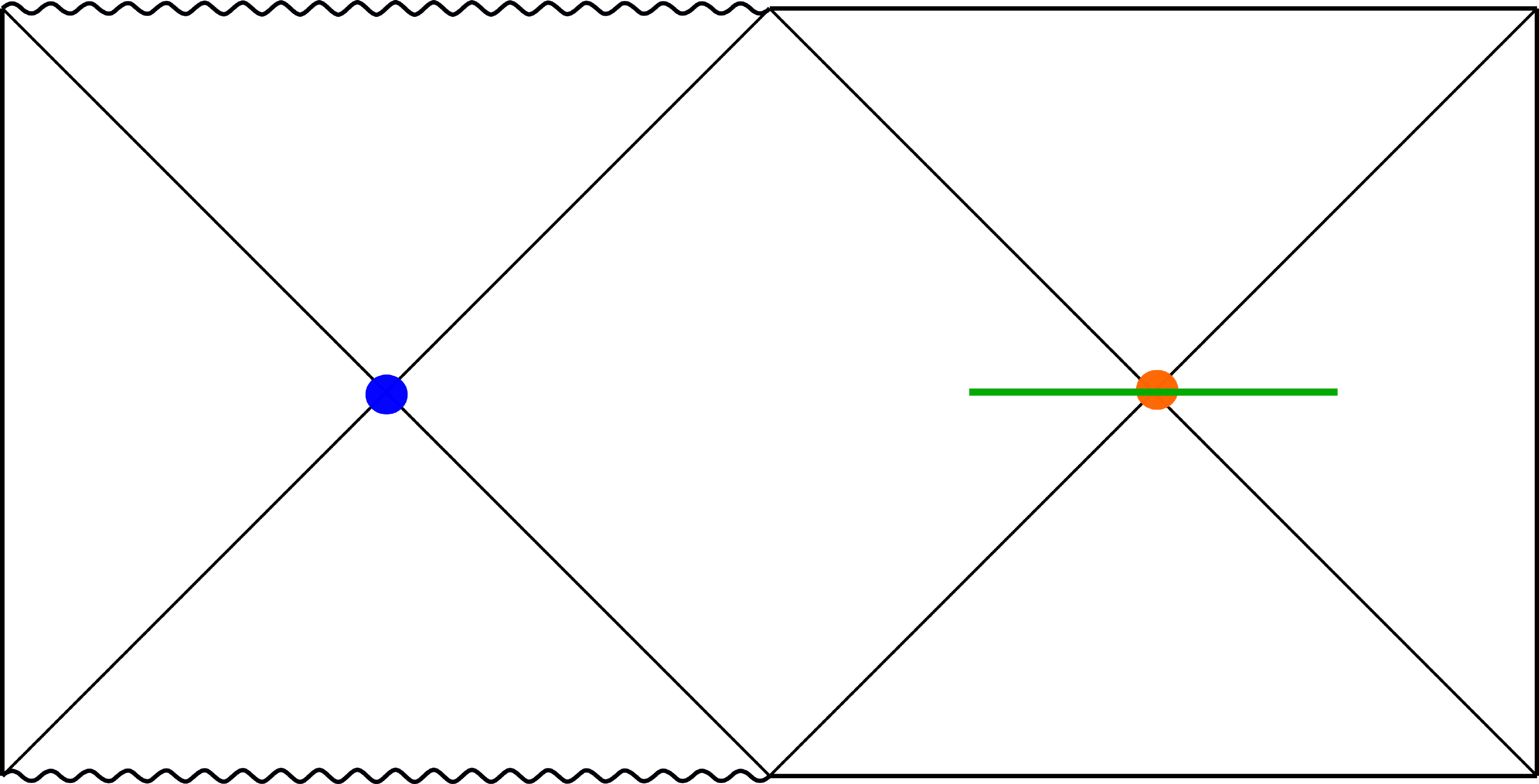}
\hspace{2 cm}
\includegraphics[width=6cm]{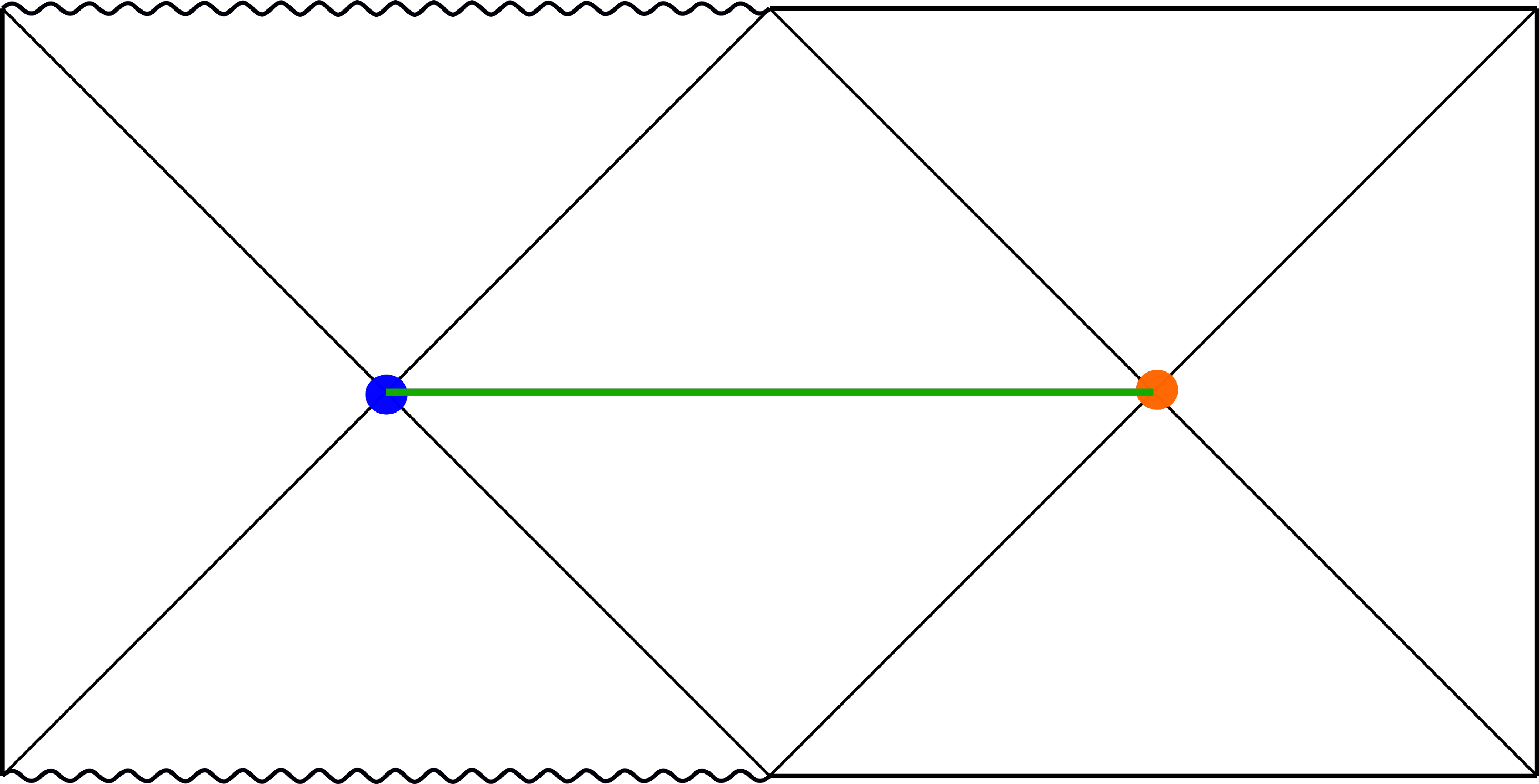}
\caption{\small{Two types of islands $C$ in the geometry without backreaction \eqref{eq:generalsourceless}, with $\zeta=0$. Instead of the islands themselves, we draw complementary regions of these islands $\bar{C}$ on the Cauchy slice $\tau=0$ (green  lines).   Left:  The complement  $\bar{C}$ of  a type I island only contains the cosmological horizon.   Right: The complement  $\bar{C}$ of   a type II island approximately connects the black hole and cosmological horizons.}}
\label{fig:nonIsland}
\end{figure}

The first candidate solution of \eqref{eq:sourceless-island-condition-1}  for $\overline{C}$ is
$\theta_{1} = -\theta_{2}$; these complements of  type I islands are therefore intervals centered on the cosmological horizon (Fig.~\ref{fig:nonIsland}).  
At low temperatures, we can approximate \eqref{eq:sourceless-island-condition-2} as
\begin{equation}
    c \cot \theta_2 - \frac{c}{\theta_2} + 6\bar{\phi} \sin\theta_2 =0 \, ,
\end{equation}
an expression which is independent of $\beta$. 
We can find a critical point for $\theta_2$ by expanding around $\theta_2 \sim \pi$ and retaining terms up to $\mathcal{O}(\theta_2-\pi)$. This produces a quadratic equation that can be solved for $\theta_2$ in terms of $c$ and $\bar{\phi}$.
By tuning $\bar{\phi}$, we can adjust the size of this island in order to make our low temperature approximation $\theta_2 \ll \beta$ accurate for arbitrary temperatures.

\subsubsection*{Type II islands}

A second candidate solution to \eqref{eq:sourceless-island-condition-1} is $\theta_2 = \pi + \theta_1$. In this case, we have an exact solution of \eqref{eq:sourceless-island-condition-2}:
\begin{equation}
    \sin \theta_1 = -\frac{\pi c}{3\bar{\phi}\beta} \coth \frac{\pi^2}{\beta} .
\end{equation}
As long as the right hand has magnitude less than 1, there will be a type II island of length $\pi$, and by tuning the parameters in our theory, we can move this island around the Cauchy slice.
At low temperatures $\beta \to \infty$, there is a solution where
\begin{equation}
    \sin \theta_1 = -\frac{c}{3\pi \bar{\phi}} .
\end{equation}

The right endpoint of the island is positioned to the right of the cosmological horizon on the Penrose diagram, where $\sin \theta_1 < 0$.
Similarly, the left endpoint is positioned to the right of the black hole horizon, and both of these endpoints approach the respective horizons for $c/\bar{\phi} \ll 1$.

At high temperatures, the interval endpoints satisfy
\begin{equation}
    \sin \theta_1 = -\frac{\pi c}{3\bar{\phi}\beta} , \quad \sin\theta_2 = \frac{\pi c}{3\bar{\phi}\beta} .
\end{equation}
For small $c/\bar{\phi}\beta$, the endpoints of the interval are near the cosmological horizon $\theta_{1} \sim 0$ and the black hole apparent horizon $\theta_{2} \sim \pi$ (Fig.~\ref{fig:nonIsland}).
Notice that for fixed $c$ and $\bar{\phi}$, the type II island either exists at low or high temperature but not in both regimes.

\subsubsection*{Type III islands}
The third and final candidate solution of \eqref{eq:sourceless-island-condition-1} is $\theta_2 - \pi = \pi - \theta_1$, so the two endpoints of  the complement $\bar{C}$ of a type III island  are both located symmetrically around the black hole horizon.
With the ansatz $\theta_{1} =\pi(1-x)$ and $\theta_{2} =\pi(1+x)$ for this $\bar{C}$, the generalized entropy is given by 
\be 
S_{\text{gen}}(x) = 2\phi_0  - 2 \bar{\phi} \cos \pi x+ \f{c}{3} \log \left[\frac{\beta}{\pi} \sinh \f{2\pi^{2}x}{\beta}   \right]-\f{c}{3} \log \left[ 2 \sin \pi x \right], \label{eq:genenbh}
\ee
and the equation for the critical point 
\eqref{eq:sourceless-island-condition-2} takes the form
\begin{equation}
    c \cot \pi x - \frac{2\pi c}{\beta} \coth \frac{2\pi^2 x}{\beta} - 6\bar{\phi} \sin \pi x = 0.
\end{equation}
There is  no solution except $x=0$ for this equation. This is because,  
as we increase the size of the interval $x$, both the dilaton and CFT field contributions to the generalized entropy are strictly increasing. As a result,  this function  \eqref{eq:genenbh} does not have a critical point except at $x=0$, and is monotonically increasing when $x > 0$. Therefore, at high temperatures we see that the complements $\overline{C}$ of  type III islands seem to disappear. Equivalently, the type III islands seem to occupy the entire Cauchy slice $\tau=0$. However, as we will see below, there is an important global subtlety in this reasoning which we will have to resolve to correctly recover the effects of entanglement monogamy.

\subsection{Islands in the backreacted solution}\label{sec:islands-backreacted}

Now we would like to take the effect of backreaction into account. The dilaton profile of the backreacted geometry is given by  \eqref{eq:totaldil2}. 
As we are now performing an honest entropy calculation in a legitimate semiclassical solution, we will specify all candidate entropies.
The first is the no-island phase thermal entropy\footnote{We are mostly interested in the high temperature limit, so we are not concerned with factors that are subleading in $\beta$ (for example, $\mathcal{O}(\log \beta)$ terms) which correct the leading $c/\beta$ behavior as $\beta \to 0$.}
\begin{equation}
    S_\beta(B) = \frac{c}{3} \log \left[ \frac{\beta}{\pi} \sinh \frac{2\pi^2}{\beta} \right] ,
\end{equation}
and the second is the island phase generalized entropy $S_{\text{gen}}[\overline{C}]$
\begin{equation}
    S_{\text{gen}}(\tau_1,\theta_1,\tau_2,\theta_2) = \phi(\tau_1,\theta_1) + \phi(\tau_2,\theta_2) + S_\beta[\overline{C}] - S_{\text{vac}}[\overline{C}] ,
\end{equation}
where we have made use of \eqref{eq:totaldil2}, \eqref{eq:thermal-cft-entropy-subregion}, and \eqref{eq:vacuum-cft-entropy-subregion}, and we must minimize over all possible islands which extremize the generalized entropy.
We will see that again there are several types of islands to consider. 
Analytically, we will be mostly interested in the high temperature limit $\beta \rightarrow 0$, but a numerical minimization would also be of interest.

\subsubsection*{Type II islands}

We now consider type II islands, which  connect the black hole and cosmological horizons. 

In the backreacted solution, we will find that type II islands are qualitatively similar to  those in the previous subsection, but support an additional solution connecting the black hole apparent horizon to a local dilaton maximum. However, the additional solution always has higher generalized entropy, and so will not be relevant.

In the high temperature limit, we expect that the size of $\overline{C}$ (as measured by either $\theta_2-\theta_1$ or $\tau_2-\tau_1$, since they appear in simple linear combinations in $S_\beta$ and $S_{\text{vac}}$) is much larger than $\beta$, so the generalized entropy we should extremize is approximated by
\be 
\begin{split}
S_{{\rm gen}} (\tau_{1},\theta_{1},\tau_{2},\theta_{2}) & =   \phi(\tau_{1},\theta_{1})+\phi(\tau_{2},\theta_{2}) +\f{\pi c}{3\beta} (\theta_{2} -\theta_{1}) + \frac{c}{3} \log \frac{\beta}{\pi} \\
& \hspace{.1cm} - \frac{c}{6} \log \left[ 2 \sin \frac{\theta_2-\theta_1+\tau_2-\tau_1}{2}\right]- \frac{c}{6}\left[ \log 2 \sin \frac{\theta_2-\theta_1-\tau_2+\tau_1}{2} \right].
\end{split}
\ee
In this limit, the conditions $\partial_{\tau_{1}}S_{{\rm gen}}=\partial_{\tau_{2}}S_{{\rm gen}}=0$ reduce to the $\tau$ extremization equations for the dilaton itself $\partial_{\tau_{1}}\phi=\partial_{\tau_{2}}\phi=0$.
The vacuum entropy $S_{\rm vac}$ terms do not contribute significantly since (as we will see) type II islands roughly connect the black hole and cosmological apparent horizons, which means that the interval size is such  that the derivatives of the sines are small in the second line.
The other conditions $\partial_{\theta_{1}}S_{{\rm gen}}=\partial_{\theta_{2}}S_{{\rm gen}}=0$ are  approximated by
\begin{equation}
    \frac{\bar{\phi} L}{2}\left( b + \frac{1}{b} \right) \frac{\sin \theta_1}{\cos \tau_1} = -\frac{\pi c}{3\beta} , \quad
    \frac{\bar{\phi} L}{2}\left( b + \frac{1}{b} \right) \frac{\sin \theta_2}{\cos \tau_2} = \frac{\pi c}{3\beta} .
    \label{eq:approxext}
\end{equation}
In the high temperature limit, we have $\tau_0 \to \frac{\pi}{2}$ and $b \sim 1/\beta^2$.

We see that there are two types of solutions at high temperature, when (after moving $b$ to the right side of \eqref{eq:approxext}) we must have $\sin \theta_1 \ll 1$ and $\sin \theta_2 \ll 1$.
The first solution connects the black hole apparent horizon at $(\tau_{2},\theta_2) =(0, \pi)$ to the maximum of the dilaton at $(\tau_{1}, \theta_1) =(0,0)$.
However, there is a solution with smaller generalized entropy. 
This second solution connects the cosmological apparent horizon $(\tau_1,\theta_1) = (\tau_0,0)$ with the black hole apparent horizon $(\tau_2,\theta_2) = (0,\pi)$.
The value of the generalized entropy for this island at high temperature is given by 
\begin{equation}
    S_{\text{type II}} \sim \phi(0,\pi)+ \phi(\tau_{0},0) = 2\phi_0 - \frac{2}{\pi} \bar{\phi} L b .
\end{equation}
Notice that we have found that, in this limit, type II islands have a generalized entropy which is decreasing as we increase temperature.
This follows from the fact that the black hole entropy measured by the area of its apparent horizon $\phi(0,\pi)$ is decreasing as we increase the entanglement with the auxiliary universe $A$.  As described earlier, increasing this entanglement is a cosmological analog of Hawking evaporation.

\subsection*{Type I islands}

\begin{figure}
\centering
\includegraphics[width=7cm]{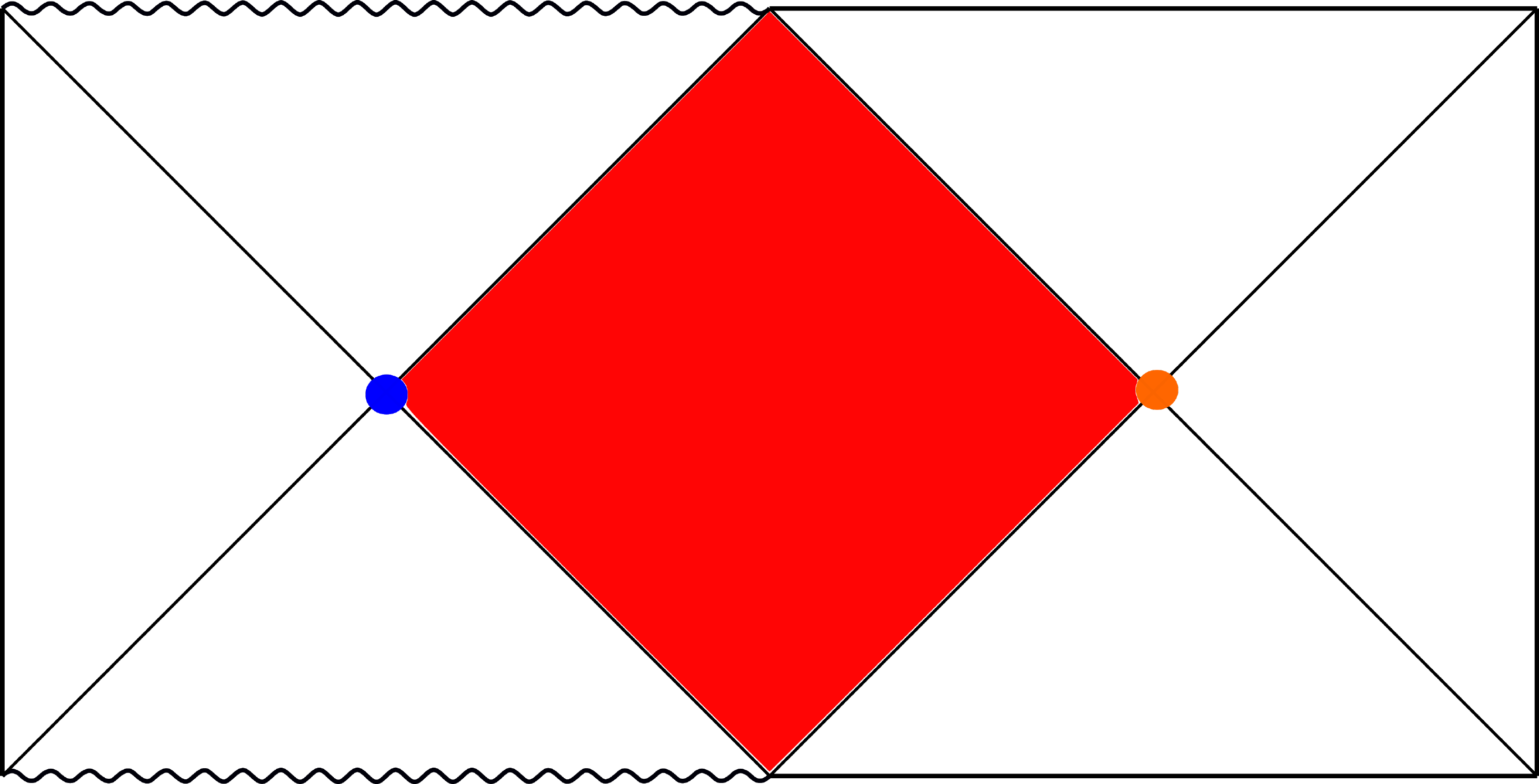}
\hspace{0.5cm}
\includegraphics[width=7cm]{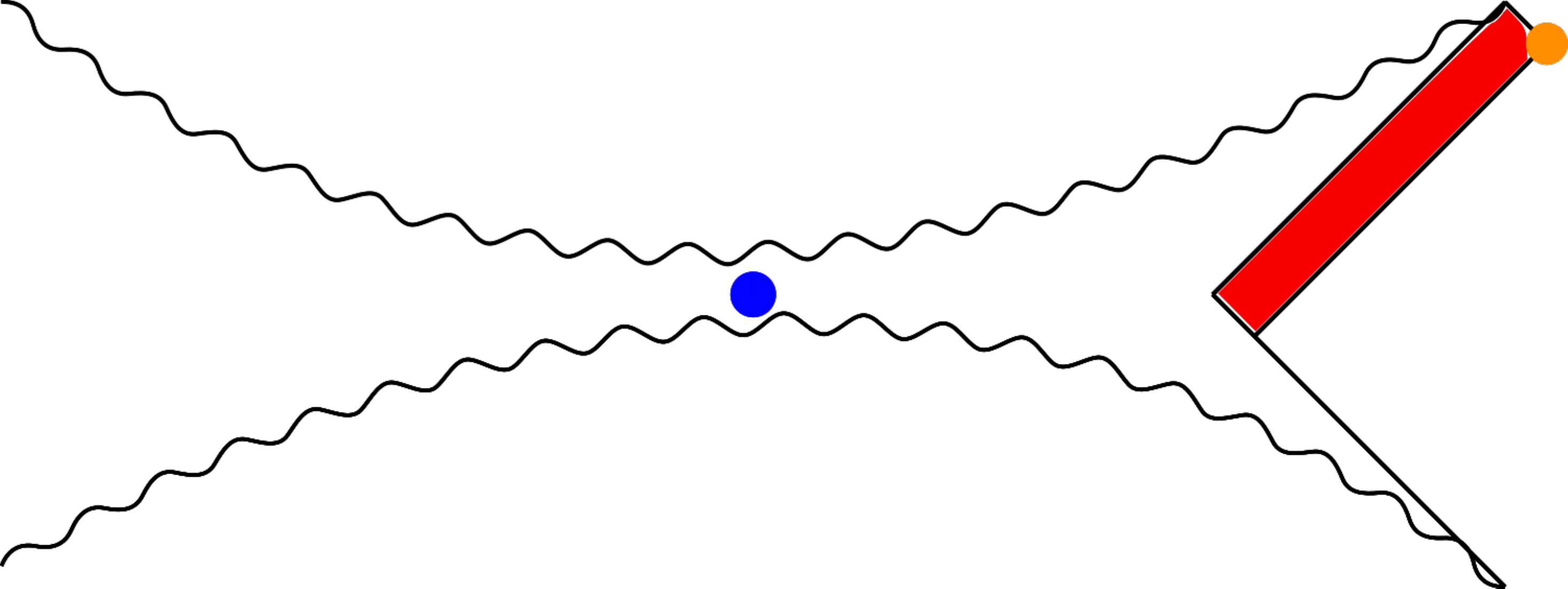}
\caption{\small{Left: The causal diamond $D(\Sigma)$ in the  spacetime without backreaction.  Right: The same  causal diamond $D(\Sigma)$ in the backreacted black hole in the high temperature limit. The causal diamond shrinks to an almost null line connecting the black hole event horizon and the cosmological event horizon.}}
\label{fig:Islanddeformed}
\end{figure}

\begin{figure}
\centering
\includegraphics[width=6cm]{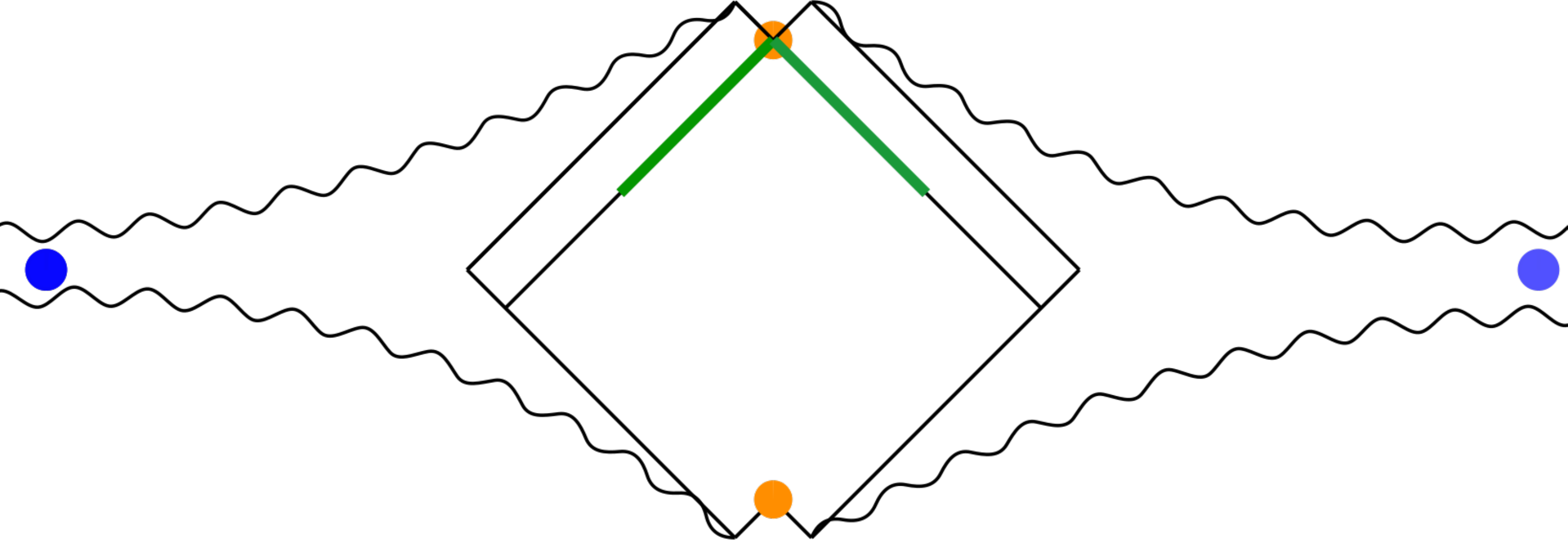}
\hspace{2 cm}
\includegraphics[width=6cm]{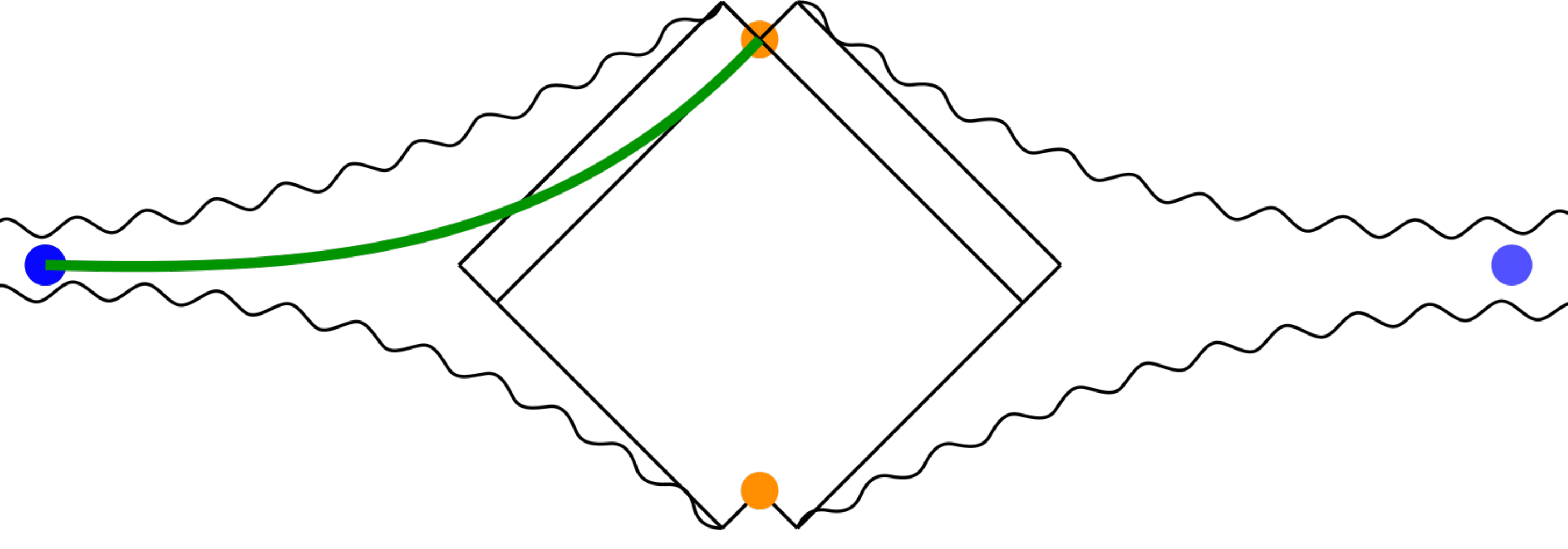}
\caption{\small{Two types of island complements, $\overline{C}$ drawn as green lines in the geometry with backreaction \eqref{eq:totaldil2} at high temperature.  Left: $\overline{C}$ of  a type I island which only contains the cosmological horizon.  Right: $\overline{C}$ of  a type II island  which approximately connects the black hole apparent horizon and cosmological apparent horizon.}}
\label{fig:Islanddef}
\end{figure}

In the high temperature limit, since $\tau_{0} \rightarrow \f{\pi}{2}$, the event horizon of the black hole at  $(\tau,\theta)=(0,\pi-\f{\tau_{0}}{2})$ and the cosmological apparent horizon $(\tau,\theta)= (\tau_{0},0) $ become null separated.   This means that a spacelike surface $\Sigma$ connecting the two horizons becomes null. Therefore, the causal diamond $D(\Sigma)$ of this surface degenerates to the null lines $\tau = \theta - \theta_+$ and $\tau = -\theta + \theta_-$, with $\tau > 0$ (see \eqref{eq:BH-event-horizon} for definitions of $\theta_\pm$).  Recall that in the sourceless solution, the complements $\bar{C}$ of  type I islands were centered on the cosmological horizon and contained within  such a spacelike slice $\Sigma$. Therefore, in the backreacted solution, it is natural to expect that in this limit the complement of  a type I island will become a part of these null lines  (see Figs.~\ref{fig:Islanddeformed} and \ref{fig:Islanddef}).

This observation motivates the following ansatz for the island,
\be
u_{1} = (\tau_{1}, \theta_{1})= ( \tau, \tau - \tau_0), \quad  u_{2} = (\tau_2, \theta_{2})= ( \tau, \tau_0 - \tau ) \, .
\ee
Furthermore, we expect that the end points satisfy $\tau_{1}, \tau_{2} \rightarrow \tau_{0}$.
By symmetry considerations, if we can find an extremum under this ansatz, we will have found an extremum of the full generalized entropy with dependence on both endpoints.
It will also be useful to record the behavior of $\tau_0$ at high temperature, obtained by expanding \eqref{eq:deftau0} around $\tau \sim \frac{\pi}{2}$ and finding the roots of the resulting quadratic equation.
\begin{equation}
    \tau_0 \sim \frac{\pi}{2} - \frac{3 \beta ^2 \bar{\phi} L }{\pi ^3 c G_N} .
\end{equation}
By plugging our ansatz into the dilaton profile \eqref{eq:totaldil2}, we find 
\begin{equation}
\begin{split}
    S_{\text{gen}}(\tau) & =   2\phi_0 + \bar{\phi}L \left( \left( b + \frac{1}{b} \right) \frac{\cos (\tau - \tau_0)}{\cos \tau} - \frac{2}{\pi} \left( b - \frac{1}{b} \right) (\tau \tan \tau + 1) \right) \\
    & \quad + \frac{c}{3} \log \left[ \frac{\beta}{\pi} \sinh \frac{\pi}{\beta} (\tau_0 - \tau) \right] - \frac{c}{3} \log \left[2\sin (\tau_0 - \tau) \right] .\label{eq:typeIgeneq}
\end{split}
\end{equation}
Numerical analysis of this function  reveals a critical point near $\tau \sim \tau_{0}  $ for $\beta \to 0$.
Therefore, in the high temperature limit we have\footnote{This can be seen by plugging   $\tau=\tau_{0}$ into the dilaton part of \eqref{eq:typeIgeneq} and using \eqref{eq:CHsourceless}. Note that when $\tau \rightarrow \tau_{0}$, the entanglement entropy part in \eqref{eq:typeIgeneq} is vanishing. }
\begin{equation}
    S_{\text{type I}} = 2\phi_0 + 2\bar{\phi} L .
\end{equation}
Note the similarity of this result to the de Sitter entropy $\phi_0 + \bar{\phi}L$.
This is consistent with the idea that the Hilbert space $\mathcal{H}_{{\rm dS}}$ of quantum gravity in de Sitter space is finite dimensional.  
Once we turn on gravity in universe $B$, we only have a finite number of states in the de Sitter Hilbert space which can become entangled with the CFT on universe $A$, and therefore in quantum gravity the entropy of $A$ must be bounded by the de Sitter entropy.

\subsubsection*{Type III islands}

Finally, let us consider the  type III islands, which connect  the two  apparent horizons of the black hole.  When there was no backreaction, we saw in Sec.~\ref{subsection:sourcelessisland}, that there is a subtlety with type III islands because of the compactness of the spatial section.  In this case, the starting point $\theta_{1}$ and the endpoint $\theta_{2}$ of the type III islands are identical, i.e. they are located at the same apparent horizon of the black hole, $\theta_{1} =\theta_{2} =\pi$.  Naively, this suggests that the Type III island occupies the whole Cauchy slice, and the that complement island therefore vanishes.  In Sec.~\ref{subsec:typeIII-compact} we will argue for an alternative interpretation where the Type III island includes the whole Cauchy slice minus a point, such that there is a non-vanishing contribution to the generalized entropy.  

In this section, to clarify matters, we will first decompactify the spatial circle of de Sitter space by passing to the universal covering space, thereby extending the black hole spacetime.  This will lead to multiple copies of the black hole (Fig.~\ref{fig:maximal}).  We will terminate the Penrose diagram on End-Of-The-World branes at $\theta =\pm \f{R}{2}$ so that the spatial section is an open interval  $-\f{ R}{2} <\theta < \f{R}{2}$.  We can think of this region as a cutoff version of the full universal cover.  In this scenario, we will find finite  type III islands.

If we write $\f{R}{2}= (2n+1) \pi +\delta$ with $0< \delta <2\pi $, then the black hole apparent horizons closest to the boundaries are at $\theta= \pm (2n+1) \pi$. For simplicity, below we choose $\delta=0$ where the leftmost and the rightmost apparent horizons are located on the boundaries. 
Our  ansatz for the type III island $C$  will be $-\frac{l}{2} \leq \theta \leq \frac{l}{2}$ on the $\tau = 0$ slice. In this case, type III islands appear between the two black hole apparent horizons which are furthest from each other in coordinate distance, i.e. between $\theta = \pm(2n+1)\pi$.

The generalized entropy functional for the complement island $\overline{C}$ is therefore given by plugging  \eqref{eq:totaldil2} and  \eqref{eq:defofb} into \eqref{eq:gen-entropy-island-phase}:
\be
\begin{split}
S_{\rm gen}(l) & = 2\phi_{0} + \bar{\phi} L \left[  \left(b+\f{1}{b} \right)\cos \f{l}{2} - \frac{2}{\pi} \left( b - \frac{1}{b} \right) \right] \\[+10pt]
& \quad +\f{c}{3} \log \left[ \f{\beta}{\pi}\sinh \f{\pi (R-l)}{\beta} \right] -\f{c}{3} \log \left[2 \sin \f{\pi l}{R} \right],  
\end{split} \label{eq:genenforIII}
\ee
and we expect $\f{l}{2} \sim (2n+1)\pi$ in the high temperature limit, because the dilaton part gives the dominant contribution.   Numerical analysis of the equation $\partial_l S_{\text{gen}}=0$  reveals a critical point near $l \approx R$ when $\frac{R}{2} = (2n+1)\pi$.

The entropy associated with this interval is dominated by the black hole entropy, $S_{\text{gen}} = 2S_{BH}$.
As $R \to \infty$, we recover the universal cover geometry.
From the above analysis, we see that no matter how large the universal cover geometry becomes, if the type III islands dominate the entropy calculation, we can reconstruct the majority of universe $B$ from universe $A$.
We simply take the island to be the largest subregion which connects two black hole apparent horizons.
The appearance of this sort of island is similar in spirit to the AdS case, where a long wormhole played the role of the island and the complement island shrunk toward the ends of the Penrose diagram \cite{WIP}.

\subsubsection*{Net result}

The actual value of the entanglement entropy of  universe $A$ is thus given by the minimizing these three contributions:

\be 
S_A = \min \{ S_{\text{no-island}}, S_{\text{type I}}, S_{\text{type II}}, S_{\text{type III}} \} .
\label{eq:minimization}
\ee
We are interested in performing this calculation for the  case where the spatial direction is an open interval $-\f{R}{2} \leq \theta \leq \f{R}{2}$. Also,  we again choose $\f{R}{2}= (2n+1)\pi$ so that the relevant apparent horizons are located on the boundaries of the interval.

Above we discussed type I and II islands in the compact case $0 \leq \theta \leq 2\pi $.  Generalizing these islands to the open interval $-\f{R}{2} \leq \theta \leq \f{R}{2}$  is straightforward.  The complement $\bar{C}$ of the type I island  is localized on a particular cosmological horizon.  The complement $\bar{C}$ of the type II island  connects a cosmological horizon and its nearest black hole  apparent horizon, which is required  in order to minimize the entanglement entropy part.   The no-island entropy, at high temperature, is also unchanged except for a factor of the universal cover cutoff $R$.
In the high temperature limit, these are given by
\be 
S_{\text{no-island}}= \f{2\pi^{2} c}{3}RT , \quad S_{\text{type I}} = 2\phi_0 + 2\bar{\phi} L , \quad S_{\text{type II}} =
2\phi_0 - \frac{\pi^3c}{6} T^2 ,
\quad 
S_{\text{type III}} = 2S_{BH} .
\label{eq:eachvalue}
\ee

\begin{figure}[t]
    \centering
    \includegraphics[scale=.6]{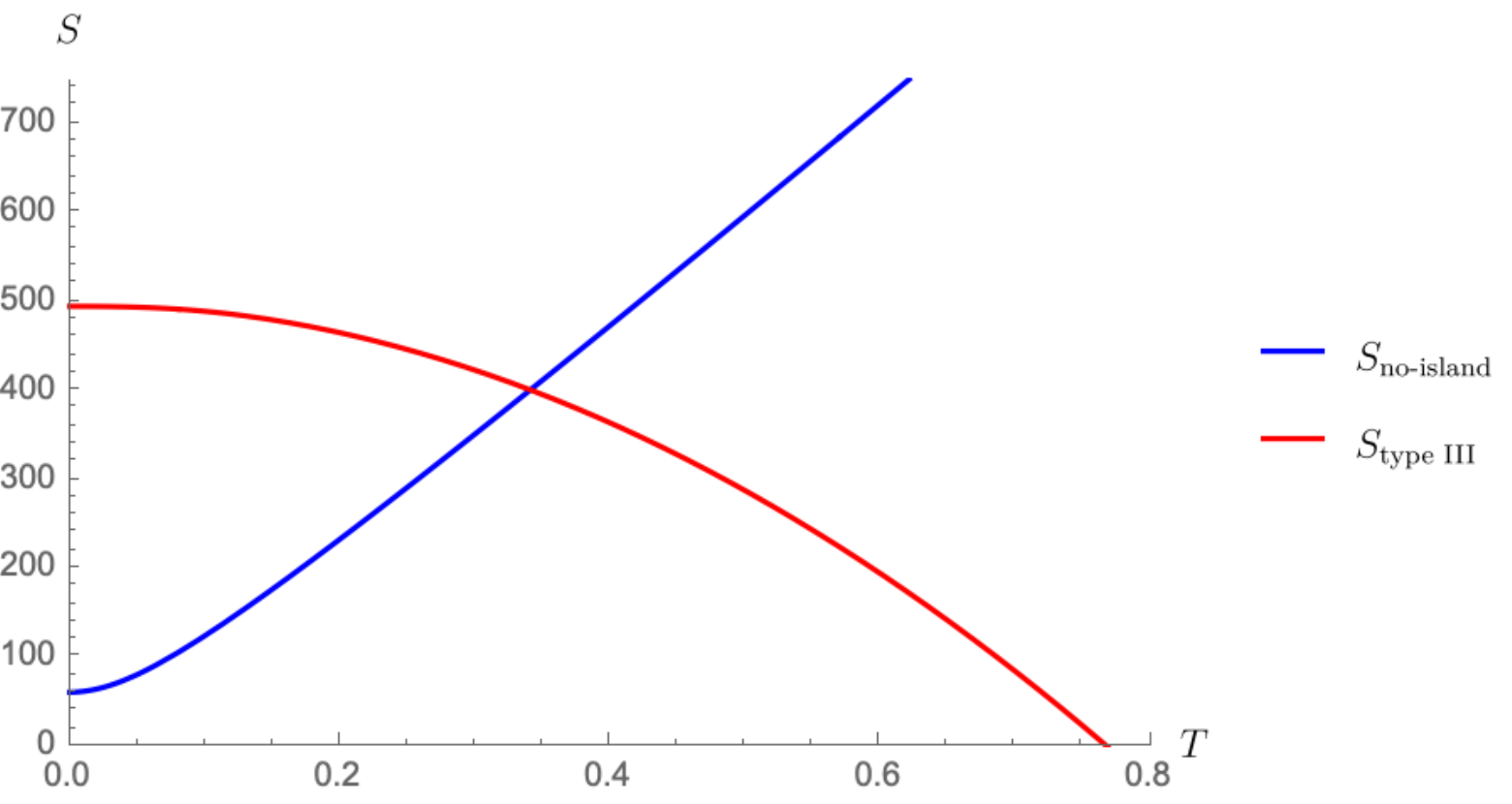}
    \caption{\small{The Page curve for the universal cover of a 2d Schwarzschild-de Sitter black hole in universe $B$.  The blue curve is the thermal entropy of CFT fields on $A$, and the red curve is the entropy of a type III island.}}
    \label{fig:ds-page}
\end{figure}

Since we  are choosing $ \f{R}{2}= (2n+1)\pi$, the end points of the type III islands get close to the boundaries. This  means that the entanglement entropy part $S_{\beta}[\overline{C}] -S_{{\rm vac}}[\overline{C}]$ in \eqref{eq:genenforIII} is vanishing, and  the generalized entropy is  equal to twice the black hole entropy $S_{\text{type III}} =2S_{BH}$.
Rewriting \eqref{eq:BHarea} in terms of $\beta$, the black hole entropy is 
\begin{equation}
\begin{split}
    S_{BH} & = \phi_0 - \frac{K'}{2} - \frac{1}{4}\sqrt{\pi^2 {K'}^2 + 16 \bar{\phi}^2 L^2} \\
    & \approx \phi_0 - \frac{(\pi+2) c}{24} - \frac{(\pi+2)\pi^3 c}{3\beta^2} ,
\end{split}
\end{equation}
where in the first line we have used $K'$ defined in  \eqref{eq:modified} and  \eqref{eq:totaldil},   and in the second line we have recorded the high temperature behavior. From these results, it is clear that we have $ S_{\text{type I}} > S_{\text{type II}}$ in the presence of the black hole.  This make sense, because as we increase the entanglement temperature, the area of the apparent horizon decreases, but the area of the cosmological apparent horizon only changes slightly, and is bounded by a maximum at zero and infinite temperatures.
Furthermore, we have $S_{\text{type III}} < S_{\text{type II}}$, as the black hole entropy is smaller than the cosmological entropy.
Thus, we have found (Fig.~\ref{fig:ds-page})
\be
  S_{A}=\begin{cases}
    S_{\text{no-island}}, &  T\leq T_{0} ,\\
S_{\text{type III}}, &  T\geq T_{0} ,
  \end{cases}
\ee
This transition temperature $T_{0}$ is computed by equating (in the high temperature limit) the thermal entropy of CFT fields on $A$ to the black hole entropy, which is the value of the dilaton at the black hole apparent horizon $(\tau,\theta) = (0,\pi)$. When $\phi_{0}/c \gg R^{2} $, we get
\begin{equation}
    T_0 \approx \sqrt{\frac{3\phi_0}{(\pi+2)\pi^3 c}} - \frac{R}{(\pi+2)\pi}. 
\end{equation}
When $ R^{2} \gg \phi_{0}/c $, we get
\be
T_0 \approx \f{3\phi_{0}}{2\pi^{2}c R}.
\ee
 
We can compare $T_0$ to the critical temperature discussed in Sec.~\ref{sec:penrose}, where we lose classical control over the solution:
\begin{equation}
    T_{\text{crit}} \approx \sqrt{\frac{3\phi_0}{(\pi+2)\pi^3 c}} > T_0.
\end{equation}
So, we can observe the Page transition to type III island dominance before losing semiclassical control of the solution for any value of $R>0$.\footnote{Indeed, this fact is implied for arbitrary (physical) values of the parameters $\phi_0$, $c$, and $R$, effectively by the intermediate value theorem.  Since $S_{BH}$ is decreasing quadratically, it must intersect the linearly increasing $S_{\text{no-island}}$ prior to reaching zero.}
Furthermore, the Page curve actually turns around at $T_0$ instead of saturating, since $S_{BH}$ is decreasing with $1/\beta$.
This decrease of the entropy after the  temperature $T_{0}$ is the analog in our setup of the Page behavior of an evaporating black hole.  Indeed, black holes in de Sitter space can evaporate, unlike large black holes in AdS.
Thus it is natural that the entropy of de Sitter black hole that is entangled with a radiation system of increasing size decreases instead of approaching a constant value.

Below we can find a similar conclusion for the compact black hole, subject to certain assumptions about the island formula which we now discuss.

\subsection{Type III islands exist when the spatial direction is compact}\label{subsec:typeIII-compact}

We have seen that when the spacetime is put on an open interval $-\f{R}{2} < \theta < \f{R}{2}$ (or its complete extension $R \rightarrow \infty$), the type III island gives the dominant contribution to the entropy  \eqref{eq:minimization} in  the high temperature limit $\beta \rightarrow 0$.  
This island connects two apparent horizons located at the boundaries of the spacetime $\theta= \pm \f{R}{2}$. We included the dilaton values $\phi (\tau=0, \theta=\pm \f{R}{2}) $ at these boundaries in the generalized entropy $S_{\text{type III}}$ \eqref{eq:eachvalue} 
because these two points are  distinct.
Including these type III islands, we get the Page curve for the entropy (Fig.~\ref{fig:ds-page}) which is decreasing when the temperature is larger than $T_{0}$.
This is consistent with the fact that two dimensional de Sitter black holes evaporate to an empty de Sitter space \cite{Bousso:1997wi, Nojiri:1998ue,Nojiri:1998ph}. 

Suppose we now identify the two boundaries of the $-\f{R}{2} < \theta < \f{R}{2}$  interval to get a compact spatial circle $S^{1}$. In this case, the two apparent horizons at $\theta= \pm \f{R}{2}$  are identified,  implying that the region $\overline{C}$ on which we compute the generalized entropy \eqref{eq:gen-entropy-island-phase} is a point.  Naively, this would lead us to  conclude that the contribution of type III island is vanishing,  $S_{\text{type III}}=0$, because the boundary of a point is an empty set.\footnote{This is exactly the situation we encountered in the discussion of the type III island in Sec.~\ref{subsection:sourcelessisland} where we restricted ourselves to the window $0 \leq \theta \leq 2\pi$ and found the absence of these islands.} Since in this case $S_{\text{type III}}$ is always smaller than $S_{{\rm no-island}}$, from  \eqref{eq:minimization}, we would appear to get $S_{A}=0$.

There are two ways to interpret this.  A first possibility is that  the above discussion is correct, and we always have $S_{A}=0$ regardless of the CFT temperature. This implies that a state in a closed universe can never become entangled with a state in an auxiliary system (in our case the universe $A$). This is the interpretation considered in  \cite{Almheiri:2019hni}.   We offer an alternative interpretation: we should regard the point-like complement island as the limit of a sequence of intervals, and hence, similar to the open interval cases, we should include (two times) the value of the dilaton at the apparent horizon.    In other words, the limit of the generalized entropy as  the type III island approaches the whole Cauchy slice is in fact $2S_{BH}$, rather than zero; assigning a zero entropy to this limit would be discontinuous.   As discussed earlier in this paper, the latter interpretation is consistent with both (a) the understanding of de Sitter space as having a Hilbert space with dimension of $O(e^{1/G_N})$, and (b) the Page curve for de Sitter black holes.  It seems that there is a subtle global issue in the interpretation of the island formula when applied to closed universes.

To reiterate, we propose that the Cauchy slice island should regarded as a ``maximal interval'',  i.e. the full  Cauchy slice minus a point, motivated by the requirement the entropy should reproduce the smooth limit of a single interval island.  Of course, in any field theory a limit that removes endpoints of intervals may not give a continuous limit for the entropy due the change in the number of UV divergences associated to the endpoints.  But here we have been dealing with manifestly finite, renormalized quantities, so this argument concerning divergences may not apply.  Indeed, in a 2d CFT there is a general argument that suggests the renormalized entropy is smooth: the OPE of a twist operator with an anti-twist operator starts with the identity, and this channel dominates as the endpoints of two different intervals approach each other.
The leading order answer for the renormalized entropy, then, is simply the answer for the configuration where the two intervals are joined.

\subsection{Cosmological islands}

Finally, we consider pure de Sitter space, without a black hole horizon but with a cosmological horizon. 
This discussion is intended to be speculative, and we leave  details to future work.

It is instructive to start from the spacetime without backreaction.   We want to construct a spacetime with a single extremum of the dilaton, and thus only a cosmological horizon.  We can try to do this by identifying $\theta=\pi/2 $ and $\theta=-\pi/2$ in Fig. \ref{fig:causalsh}.  
We introduce the identification to demand that the resulting configuration is free from the singularity which originates from the black hole; otherwise, the initial value problem is not well defined in the cosmological model. Another way to see this is that, after the coordinate transformation to the static patch, 
\be
r= \f{\cos \theta}{\cos \tau}, \quad \tanh t= \f{\sin \tau}{\sin \theta},
\ee
the metric as well as dilation have the form \eqref{eq:static}. In these coordinates, it is clear that pure de Sitter space is constructed by identifying   $r=0$ of the left and right wedges of this region.
Of course, if we wished, we could employ a similar trick as we did in the universal cover discussion, where we focused on a cutoff version of the spacetime.
In the pure de Sitter case, this cutoff would be the region of $\theta$ for which the dilaton is positively diverging as $\tau \to \frac{\pi}{2}$, i.e. the past lightcone of the future asymptotic region.

Now we consider the spacetime with backreaction, so the dilaton is given by \eqref{eq:totaldil2}. In doing so we first focus on the causal diamond $D(\Sigma)$ of a spatial slice $\Sigma$ connecting the cosmological event horizon and one of the event horizons of the black hole (the right panel of Fig.~\ref{fig:Islanddeformed}).  Again we can introduce static coordinates $(t,r)$ on $D(\Sigma)$ by the coordinate transformation
\be 
\tanh t=\f{b_{+} \sin \tau -b_{-}\cos \theta}{\sin \theta}, \quad  r=b_{+}  \f{\cos \theta}{\cos \tau} -b_{-}\tan \tau, \quad b_\pm = \f{1}{2} \left(b \pm \f{1}{b} \right) .
\ee
This is because the causal diamond of the sourceless solution $\Phi_{0}$ and $D(\Sigma)$ are related by an $SL(2,R)$ boost symmetry.
One can easily check that the corners of $D(\Sigma)$ are correctly mapped to that of the undeformed causal diamond (the left panel of Fig.~\ref{fig:Islanddeformed}).
These corners are the cosmological horizon, the black hole horizon, and the two points where the past and future singularities meet the past and future asymptotic regions.

We then construct a new geometry by identifying two lines, i.e. $r=0$ in $D(\Sigma)$ and the analogous $r=0$ line in the right diamond (See Fig.~\ref{fig:Islanddef}).
Since in the new coordinates $(t,r)$ the metric is still given by \eqref{eq:static}, the backreaction as well as the new identification at $r=0$ do not change the CFT partition function, and therefore $S_{\text{no-island}}$ is unchanged as well.
In this geometry, it is clear that only type II islands exist, so the entropy curve is given by
\be
  S_{A}=\begin{cases}
    S_{\text{no-island}},\quad  T\leq \f{\phi_{0} +\bar{\phi}}{c}\\
S_{\text{type II}}=2(\phi_{0} +\bar{\phi}L),\quad  T\geq \f{\phi_{0} +\bar{\phi}}{c} .
  \end{cases}
\ee
We do not need to worry about losing semiclassical control because we have restricted our attention to the portion of the geometry which is well-separated from the approaching singularity and black hole apparent horizon.   

We have not checked the dynamical consistency of these solutions and leave this for future work.

\section{Discussion}\label{sec:discussion}

We considered black holes in 2d de Sitter JT gravity coupled  to  a  CFT, and entangled with matter in a disjoint non-gravitating  universe.  We showed that the entanglement entropy of the matter respects monogamy as strength of entanglement is increased, provided it is computed using the``island formula'' adapted to this context \cite{WIP}. We also showed the entropy formula is consistent with the interpretation of de Sitter space as having a Hilbert space with a finite dimension of $O(e^{1/G_N})$.  In a decompactified version of the de Sitter geometry, these results followed from the competition between the effective field theory entropy and the area of the boundaries of extremal islands in the gravitating geometry.   In the compact de Sitter geometry there was an interesting subtlety:  we argued that the relevant island covered the entire Cauchy slice except for a point at the apparent black hole horizon.  Equivalently, the complement island could be regarded as a Planck-sized interval surrounding the apparent horizon.

The results in the compact case touch upon a subtle issue concerning the island formula. It would appear that a closed universe will always admit an island that occupies the entire Cauchy slice, and which therefore purifies any auxiliary entangling system while also having zero boundary area.  Thus, at first glance, the island formula seems to be saying that the entropy of an auxiliary system entangled with a closed universe must be zero \cite{Almheiri:2019hni}.  But this conclusion poses several conceptual difficulties.  For example, in the $G_N \to 0$ limit where we turn off gravity it is certainly possible to entangle disjoint systems.  It would be very surprising if this entanglement vanishes for even an infinitesimal coupling.  There would be no tension if closed universes necessarily have one dimensional Hilbert spaces.  However, longstanding arguments suggest that de Sitter spaces (like the ones we study, and like the one we might be living in) have a finite dimensional Hilbert space controlled by the cosmological constant and non-perturbatively large in the Newton constant.

We proposed an interpretation that avoids these difficulties while giving a consistent semiclassical account of the properties of quantum entanglement:  the Cauchy slice island is really a maximal interval, i.e. a Cauchy slice minus a point.     The rough justification of this is the observation that entropy function evaluated on the the full Cauchy slice island does not reproduce the smooth limit of a single interval island, while our interpretation does.   Of course, once we consider multiple islands in the gravitating region, we already do not have smooth limits in the different topological sectors, e.g. the limit of a two-island entropy as two of the four endpoints approach each other does not reproduce the one-island entropy with two endpoints.\footnote{We thank Simon Ross for pointing this out to us.}

The universal cover of the de Sitter black hole has a non-compact Cauchy slice, so these issues would seem to be irrelevant in that case. However, from studies of maximally extended black holes in AdS/CFT \cite{Engelhardt:2015gla,Balasubramanian:2019qwk}, which is morally (but not precisely) similar to passing to the universal cover, it has become clear that working with the maximal extension does not correspond to including additional degrees of freedom or Hilbert space factors. Instead, correlation functions of operators placed in different patches of the extension are related to correlators in a single copy of the geometry after a certain analytic continuation procedure \cite{Balasubramanian:2019qwk}. Therefore, we would expect roughly the same behavior for a single copy of the geometry as the maximal extension or universal cover, from the microscopic perspective, if there is a genuine quantum mechanical system which describes de Sitter quantum gravity (as there famously is in AdS).   Thus the fact that we recover the Page behavior in the universal cover of the de Sitter black hole suggests that we should also recover it in the compact case.

In the compact case, we found one type of  extremal island which has generalized entropy equal to twice the de Sitter entropy in the high temperature limit.   This means that no matter what effective field theory entropy the CFT fields on universe $A$ have, the true quantum gravitational entropy is bounded by a constant related to the de Sitter entropy. This is in agreement with  expectations about the Hilbert space of de Sitter quantum gravity, and aligns with recent results \cite{Chen:2020tes}. Note that this upper bound on the entropy of universe $A$ is effectively invisible at the semiclassical level, because we lose control of the semiclassical solution long before it would be relevant for the entropy calculation. Of course, we could choose to ignore this issue and focus only on the coordinate range covering the asymptotic region, which is well-separated from the region where the singularities are meeting at the black hole apparent horizon.  Perhaps the fact that this upper bound is hidden semiclassically in a subtle way is related to the difficulty of sensing the finite de Sitter Hilbert space from a matrix model point of view \cite{Maldacena:2019cbz,Cotler:2019dcj,Cotler:2019nbi}.

\subsection*{Acknowledgments}

We thank Norihiro Iizuka, Yuki Miyashita, Simon F.~Ross, Masaki Shigemori, and Tadashi Takayanagi for useful discussions, and especially Kotaro Tamaoka for initial collaboration.
VB and AK were supported in part by the Simons Foundation through the It From Qubit Collaboration (Grant No.~38559) and DOE grant DE-SC0013528.  VB was also supported by the DOE grants FG02-05ER-41367 and QuantISED grant DE-SC0020360. TU was supported by JSPS Grant-in-Aid for Young Scientists  19K14716.   VB also thanks the Aspen Center for Physics, which is supported by National Science Foundation grant PHY-1607611.

\appendix

\section{Embedding space}
 
In the body of the paper, we used the coordinate transformation \eqref{eq:coordtrans} between  static and global coordinates.
In this appendix, we describe its derivation.
It is convenient to use the embedding space formalism. 
Two dimensional de Sitter space can be obtained by starting with a hyperboloid  
\be 
-X_{0}^{2} +X_{1}^{2} + X_{2}^{2}=1 ,
\ee
in $\mathbb{R}^{1,2}$ with the embedding space metric
\be 
ds^{2}= \eta^{AB} dX_A dX_B = -dX_{0}^{2} +dX_{1}^{2} + dX_{2}^{2} .
\ee
Static coordinates $(t,r)$ are produced by pulling back this metric via
\be 
X_{0}=\s{1-r^{2}}\sinh t, \quad X_{1}=\s{1-r^{2}}\cosh t, \quad X_{2}=r.
\label{eq:static2}
\ee
On the other hand, global coordinates $(\tau,\theta)$ may be defined with the embedding
\be 
X_{0}=\tan \tau, \quad X_{1} =\f{\sin \theta }{\cos \tau}, \quad X_{2} =\f{\cos \theta }{\cos \tau} .
\ee
The $SO(1,2)$ embedding space isometry
\be 
\left(\begin{array}{c}
    X_{0} \\
    X_1 \\
    X_{2}
  \end{array}
  \right)
  \rightarrow
  \left(\begin{array}{c}
    X_{0}' \\
    X_1' \\
    X_{2}'
  \end{array}
  \right)
  =\left(
  \begin{array}{ccc}
     b_{+}& 0 &-b_{-} \\
     0 & 1 & 0 \\
     -b_{-} & 0 & b_{+}
  \end{array}
 \right)
 \left(\begin{array}{c}
    X_{0} \\
    X_1 \\
    X_{2}
  \end{array}
  \right), \quad b_{\pm}=\f{1}{2}\left(b \pm \f{1}{b} \right) ,
\ee
leaves the hyperboloid invariant (remembering that the hypersurface definition is $X_A X^A = 1$, and the embedding space metric implies $X^0 = -X_0$).
Thus 
\be 
X_{0} =b_{+} \tan \tau -b_{-} \f{\cos \theta }{\cos \tau}, \quad X_{1}=\f{\sin \theta }{\cos \tau}, \quad X_{2}=-b_{-}\tan \tau +b_{+}\f{\cos \theta }{\cos \tau} ,
\ee
defines another global coordinate system for the embedded dS$_2$.
By equating this with \eqref{eq:static2}, we get the coordinate transformation \eqref{eq:coordtrans}.

\bibliographystyle{JHEP}
\bibliography{deSitter}

\end{document}